\documentstyle[preprint,prb,aps]{revtex}
\begin{document}
\draft
\title{Density of states of a layered S/N d--wave superconductor}
\author{W. A. Atkinson, J. P. Carbotte}
\address{Department of Physics and Astronomy, McMaster University, \\
Hamilton, Ontario, Canada L8S 4M1}
\date{\today}
\maketitle
\begin{abstract}
We calculate the density of states of a layered superconductor
in which there are two layers per unit cell.  One of the layers
contains a d--wave pairing interaction while the other is a normal
metal.  The goal of this article is to understand how the d--wave
behaviour of the system is modified by the coupling between the
layer--types.  This coupling takes the form of coherent, single particle
tunneling along the c--axis.  We find that there are two physically
different limits of behaviour, which depend on the relative locations
of the Fermi surfaces of the two layer--types.  We also discuss the
interference between the interlayer coupling and pairing interaction
and we find that this interference leads to features in the density
of states.
\end{abstract}
\narrowtext
%
%%%
\section{Introduction}

One of the most interesting developments in the study of high--temperature
(H$T_c$) superconductors has been the idea that the gap is not isotropic
within the copper--oxide planes.
There is considerable experimental evidence that this is the case;
angle resolved photoemission experiments \cite{Dessau}, for example,
have directly observed an anisotropic gap in the {\em ab}--plane spectrum of
Bi$_2$Sr$_2$CaCu$_2$O$_{8+\delta}$.  Furthermore, there
have been a number of recent experiments \cite{Raman} in which the
behaviour of the system can be explained by gaps which have at least
one node on the Fermi surface.  There have also been a number of
controversial tunneling experiments (for example Wollman {\em et al}
\cite{vanH}) in which the sign of the
gap has been shown to change across the nodes.  These experimental
results have been taken as strong support for a gap with d--wave
symmetry within the copper--oxide planes.

The d--wave symmetry of the gap is an integral part of a number of
theories of superconductivity, such as RVB \cite{Kotliar,Lee} and
spin--fluctuation \cite{Pines,Annett,Lenck} models.
There are also
a large number of BCS--like calculations
\cite{Zhou,Fedro,Carbotte,Bulut,Arberg,Kim}
(of which we have only mentioned a few) in which the ansatz is made
that the gap has a d--wave structure.
The one feature which is common to all of these calculations is that
they are 2--dimensional, and it is assumed that the effects of
the third dimension will
be weakly perturbative.  This assumption reflects the belief that the
superconductivity of the H$T_c$ materials occurs within the copper--oxide
planes.

The goal of this work is to consider the effect that the third dimension
will have on the properties of an otherwise d--wave system.  In particular,
we will calculate the superconducting density of states (DOS) of a layered
S/N system.  In this model, there are two layers in the unit cell.
One of the layers (S) contains a d--wave pairing interaction which
drives the superconducting transition.  The second layer (N) is intrinsically
normal in the sense that it contains no pairing interaction, although it
still becomes superconducting because of its proximity to the S layer.
Of the H$T_c$ materials,
this model has the most in common with YBa$_2$Cu$_3$O$_7$, where it is
considered that the copper--oxygen planes contain a superconducting
pairing mechanism and the copper--oxygen chains are intrinsically normal.

Our particular version of the layered S/N model is based on one
studied previously \cite{Abrikosov,Buzdin,Bulaevskii,Tachiki}, although
the focus of these authors was somewhat different than here.
In all cases, the authors considered the gap of the isolated
superconducting
plane to be isotropic, and the discussion focussed on the anisotropy
introduced by the interplane coupling.
Abrikosov and Klemm \cite{Abrikosov}
were interested in how the interplane coupling affects the density
of states, the isotropy of the Raman spectrum and the behaviour
of $T_c$ and the gap.  Buzdin {\em et al} \cite{Buzdin} calculated the
density of states, while Bulaevskii \cite{Bulaevskii} considered various
properties of a layered S/N model in the limits of very weak and very strong
interlayer coupling.  Tachiki {\em et al.} have calculated a number of
observable quantities (tunneling conductance, optical conductivity,
knight shift and nuclear magnetic relaxation rate) for a more complicated
model in which there are five layers per unit cell.
In this paper, the pairing in the S layer is chosen to have a d$_{x^2-y^2}$
symmetry and we are interested in how the interplane coupling will
change the density of states from the 2--dimensional case \cite{Zhou,Fedro}.

The layout of the paper is as follows:  In section \ref{sec-DOSI} we
introduce a model for a layered S/N superconductor and derive a formula
for the density of states.
In section \ref{sec-DOSII} we calculate the DOS for different choices
of model parameters.  The discussion is broken up into two physically
different limits.  In one case, the Fermi surfaces of the two bands
(which result from the two layer types) are far apart in the Brillouin
zone, while in the second case they coincide.  The difference between
these two limits is that, in the second case, there are important
(non--perturbative) interference effects between the mean field and the
interplane coupling.  In section \ref{sec-conc} we present
a summary and discussion of our results.

%%%
\section{The Model}
\label{sec-DOSI}
In this section we introduce a model for a layered superconductor, which
is based on one proposed by Abrikosov \cite{Abrikosov}.  In this model
we consider two types of metallic layers, stacked alternately one above the
other in the z direction.  These layers form two sublattices which we will
assume are weakly coupled to each other.  There is a BCS--like pairing
potential localized about the layers of the first sublattice which is
responsible for the superconductivity of the sample.
The Hamiltonian for such a system can be written
\begin{eqnarray}
  \label{0}
  {\bf H} & = & \sum_\sigma \int d^3r \, \Psi^\dagger_\sigma ({\bf r})
  (H_0 + V_1 +V_2) \Psi_\sigma({\bf r})  \nonumber \\
    &+& \sum_{\sigma \sigma^\prime} \int d^3r \, d^3r^\prime \,
  \Psi^\dagger_\sigma ({\bf r}) \Psi^\dagger_{\sigma ^\prime}({\bf r}^\prime)
\,
  I_1({\bf r},{\bf r}^\prime) \, \Psi_{\sigma^\prime} ({\bf r}^\prime)
  \Psi_\sigma ({\bf r})
\end{eqnarray}
where $\Psi_\sigma$ is the field operator of spin $\sigma$.  The operator
$H_0$ is the free particle Hamiltonian, $p^2/2m$, while $V_1$ and $V_2$ are
the potentials associated with each of the sublattices.  The BCS--like
pairing potential is
\[
  I_1({\bf r},{\bf r}^\prime) = -1/2 \sum_{Z} V(x,y,x^\prime,y^\prime)
  \delta (z-Z) \delta (z^\prime - Z),
\]
where $Z$ are the $z$--values of the planes in the first sublattice.

If the electrons are tightly bound to their planes, then the eigenstates
of $H_0 + V_1 + V_2$ are well approximated by eigenstates of the sublattice
Hamiltonians $H_0 + V_1$ and $H_0 + V_2$ (which are assumed known).
Eigenstates of $H_0 + V_1$ and $H_0 + V_2$ are not orthogonal to each other,
but may be orthogonalised to form a new set of states, $\phi_{i {\bf k}}
({\bf r}), i=1,2$ , which are no longer eigenstates of the sublattice
Hamiltonians but which satisfy $\int d^3r \, \phi^\ast_{1 {\bf k}} ({\bf r})
\phi_{2 {\bf k}^\prime} ({\bf r}) = 0$.  The orthogonalisation procedure will
conserve the Bloch--like properties of the original eigenstates.  If
we define
\[
  {\bf a}_{i {\bf k} \sigma} =  \int d^3r \,\phi^\ast_{i {\bf k}}({\bf r})
  \Psi_\sigma ({\bf r})
\]
then the Hamiltonian may be written
\begin{eqnarray}
  \label{1}
   {\bf H} - {\bf N}\mu & = & \sum_{{\bf k} \sigma} \left [ {\bf a}^\dagger
  _{1 {\bf k}
  \sigma} {\bf a}_{1 {\bf k} \sigma} \xi_1({\bf k})
  + {\bf a}^\dagger_{2 {\bf k} \sigma} {\bf a}_{2 {\bf k} \sigma}
  \xi_2({\bf k}) \right ] \nonumber \\
           & + & \sum_{{\bf k} \sigma} \left [ t({\bf k})
  {\bf a}^\dagger_{1 {\bf k} \sigma} {\bf a}_{2 {\bf k} \sigma}
  + t^\ast({\bf k}) {\bf a}^\dagger_{2 {\bf k} \sigma}{\bf a}_
  {1 {\bf k} \sigma}\right ] \\
           & - & \sum_{\bf k} \left [ \Delta_{\bf k}
  {\bf a}^\dagger_{1 {\bf k} \uparrow}{\bf a}^\dagger_{1 -{\bf k} \downarrow} +
  \Delta^\ast_{\bf k} {\bf a}_{1 -{\bf k} \downarrow}
  {\bf a}_{1 {\bf k} \uparrow}  \right ] \nonumber
\end{eqnarray}
where
\[
  \xi_j({\bf k})  =  \int d^3r \, \phi^\ast_{j {\bf k}} ({\bf r})
  (H_0 + V_1 + V_2 - \mu) \phi_{j {\bf k}} ({\bf r})
\]
\[
  t({\bf k}) =  \int d^3r \, \phi^\ast_{1 {\bf k}} ({\bf r})
  \, (H_0 + V_1 + V_2) \, \phi_{2 {\bf k}} ({\bf r})
\]
and
\begin{equation}
  \label{2}
  \Delta_{\bf k} \equiv \frac{1}{\Omega} \sum_{{\bf k} ^\prime}
  V_{{\bf k k}^\prime} \langle {\bf a}_{1 -{\bf k}^\prime \downarrow}
  {\bf a}_{1 {\bf k}^\prime \uparrow} \rangle .
\end{equation}
Here $\Omega$ is the volume of the sample and $\langle \rangle$ denotes
a thermal average.  Equation (\ref{2}) is the self--consistent equation
which follows from making a mean field approximation for the pairing
interaction in equation (\ref{0}).
The form of the potential, $V_{{\bf k k^\prime}}$,
comes from assuming a singlet pairing of electrons of opposite ${\bf k}$.  We
find that
\begin{eqnarray}
  V_{{\bf k k}^\prime} & = & \sum_{Z} \int dxdy \, dx^\prime dy^\prime
  \phi^\ast_{1 {\bf k}} (x,y,Z) \phi^\ast_{1 -{\bf k}}(x^\prime,y^\prime,Z)
  \nonumber \\
  &  \times &   V(x,y,x^\prime,y^\prime)
  \phi_{1 -{\bf k}^\prime}(x^\prime,y^\prime,Z)\phi_{1 {\bf k}^\prime}
  (x,y,Z) \nonumber,
\end{eqnarray}
which, in the limit in which the electrons are tightly bound to the
planes, has no $k_z$ or
$k_z^\prime$ dependence.  The interplane hopping term, $t$, has been
chosen to conserve ${\bf k}$, which makes sense if $V_1$ and $V_2$
have the same periodicity and we have a clean metal.

In the Nambu formalism, the Hamiltonian can be written
\begin{equation}
  \label{4a}
  {\bf H} - {\bf N}\mu = \sum_{\bf k} {\bf A}^\dagger ({\bf k})Q({\bf k}){\bf
A}
  ({\bf k}) + \mbox{const}
\end{equation}
where
\begin{equation}
  \label{5}
  {\bf A} ({\bf k}) = \left [ \begin{array}{c}
  {\bf a}_{1 {\bf k} \uparrow} \\
  {\bf a}^\dagger_{1 -{\bf k} \downarrow} \\
  {\bf a}_{2 {\bf k} \uparrow} \\
  {\bf a}^\dagger_{2 -{\bf k} \downarrow}
  \end{array} \right ]
\end{equation}
and
\begin{equation}
  \label{6}
  Q =  \left [ \begin{array}{cccc}
  \xi_1({\bf k}) & -\Delta_{\bf k} & t({\bf k}) & 0 \\
  -\Delta^\ast_{\bf k} & -\xi_1(-{\bf k}) & 0 & -t^\ast(-{\bf k}) \\
  t^\ast({\bf k}) & 0 &  \xi_2({\bf k}) & 0 \\
  0 &  -t(-{\bf k}) & 0 & -\xi_2(-{\bf k})
  \end{array} \right ] .
\end{equation}
The diagonalisation of the matrix, $Q({\bf k})$, will be simplified if we
note that the symmetries $t({\bf k}) = t^\ast(-{\bf k})$ and
$\xi_i({\bf k}) = \xi_i(-{\bf k})$ follow from time reversal symmetry.
Then the energy eigenvalues are $E_1 = E_+$, $E_2 = E_-$, $E_3 = -E_-$,
$E_4 = -E_+$, with
\begin{eqnarray}
  \label{7}
  E_\pm ^2 & = & \frac{\xi_1^2 + \xi_2^2 + \Delta_{\bf k}^2}{2} + t^2
  \nonumber \\
  & \pm &\sqrt{ \left [ \frac{\xi_1^2 - \xi_2^2 + \Delta_{\bf k}^2}{2}
  \right ]^2 + t^2 [ (\xi_1 + \xi_2)^2 + \Delta_{\bf k}^2] } \\
  \label{8}
           & = & \frac{\xi_1^2 + \xi_2^2 + \Delta_{\bf k}^2}{2} + t^2
  \nonumber \\
  & \pm & \sqrt{ \left [ \frac{\xi_1^2 + \xi_2^2 + \Delta_{\bf k}^2}{2} + t^2
  \right ]^2
  - (t^2 - \xi_1 \xi_2)^2 - (\xi_2 \Delta_{\bf k})^2 } \nonumber \\
\end{eqnarray}
(we simplify our notation by understanding $t^2$ to mean $|t|^2$ and
$\Delta_{\bf k}^2$ to mean $|\Delta_{\bf k}|^2$ throughout this paper).
The first expression for $E_{\pm}$, given by
equation (\ref{7}), is useful for describing the weakly coupled (small $t$)
limit, while the second, equation (\ref{8}), is useful for
describing $E_-$ near its minima.

The Hamiltonian now becomes
\begin{equation}
  {\bf H} - {\bf N}\mu = \sum_{\bf k} \sum_{i=1}^{4} \hat{\bf A}^\dagger_i
  ({\bf k}) \hat{\bf A}_i({\bf k}) E_i({\bf k}) + \mbox{const}
\end{equation}
where $\hat{\bf A}_i({\bf k})$ is the quasiparticle annihilation operator
associated with the $i^{\mbox{th}}$ energy band.   The operator,
$\hat{\bf A}_i({\bf k})$, is defined by
\begin{equation}
  \label{8b}
  \hat{\bf A}_i({\bf k}) = \sum_{j=1}^{4} U^\dagger_{ij} ({\bf k})
  {\bf A}_j ({\bf k}),
\end{equation}
and $U ({\bf k})$ is the $4 \times 4$ matrix which diagonalises $Q$:
$U_{ij} = U_i(E_j)$ with
\begin{eqnarray}
  \label{9}
  U_i(E_j)& = & \frac{1}{\sqrt{C}} \left [\begin{array}{c}
               (E_j - \xi_2) A \\
	       -(E_j + \xi_2) B \\
	       t A \\
	       t B \\
	       \end{array} \right ]
\end{eqnarray}
\[
  A  =  t^2 - (\Delta_{\bf k} + E_j + \xi_1)(E_j + \xi_2)
\]
\[
  B  =  t^2 - (\Delta_{\bf k}^\ast + E_j - \xi_1)(E_j - \xi_2)
\]
\[
  C =  A^2 [t^2 + (E_j - \xi_2)^2] + B^2 [t^2 + (E_j + \xi_2)^2].
\]

  We are now in a position to solve for the temperature dependence of the
gap, $\Delta_{\bf k}$.  We do this by writing equation (\ref{2}) in terms of
the quasiparticle operators, $\hat{\bf A}_j ({\bf k})$.
We get
\begin{equation}
  \label{10}
  \Delta_{\bf k} = \frac{1}{\Omega} \sum_{i=1}^4 \sum_{{\bf k}^\prime}
  V_{{\bf k k}^\prime}
  U_{1i}({\bf k}^\prime) U_{2i}({\bf k}^\prime) f(-E_i({\bf k}^\prime))
\end{equation}
where $f(x) = [1 + \exp(\beta x)]^{-1}$ is the Fermi distribution function.
Equation (\ref{10}) is a self--consistent equation which can be
solved numerically for the gap as a function of temperature.

Once the gap at $T=0$ is known, then the density of states (DOS) may be
calculated.  The density of states is
\[
  \rho(\omega) = - \frac{2}{N \pi} \int d^3r \,
  \mbox{{\bf Im}} \, G^R_{\uparrow\uparrow}(r,r,\omega)
\]
where the factor $2$ is because we are only counting spin up electrons and
N is the number of unit cells in the crystal.
$G^R_{\sigma \sigma^\prime}$ is the retarded Green's function:
\begin{eqnarray}
  G^R_{\sigma \sigma^\prime}(r,r^\prime,w) &=& \lim_{\delta\rightarrow 0^+}
  \int_{-\infty}^{\infty} dt \, e^{i(w+i\delta)t} \nonumber \\
  & & \times  \langle \{
   \Psi_\sigma (r,t),  \Psi^\dagger_{\sigma^\prime} (r^\prime,0) \}
  \rangle \theta(t)  \nonumber
\end{eqnarray}
where $\theta(t)$ is the step function and \{ \} are anticommutating
brackets.  If we write
\begin{eqnarray}
  \Psi_\uparrow (r,t) & = &  \sum_{i=1}^2 {\bf a}_{i {\bf k} \uparrow} (t)
  \phi_{i {\bf k}} (r) \nonumber \\
   & = & \sum_{i=1}^4 \left \{ U_{1i}({\bf k}) \hat{\bf A}_i({\bf k},t)
  \phi_{1 {\bf k}} (r) \right. \nonumber \\
   & & \left. + U_{3i}({\bf k}) \hat{\bf A}_i({\bf k},t)
  \phi_{2 {\bf k}} (r) \right \}
\end{eqnarray}
and
\[
  \hat{\bf A}_i({\bf k},t) = e^{-i E_i({\bf k}) t} \hat{\bf A}_i({\bf k},0)
\]
then
\begin{equation}
  \label{11}
  \rho(\omega) =  \frac{2}{N} \sum_{i=1}^4 \sum_{\bf k} \left [
  U_{1i}({\bf k})^2 + U_{3i}({\bf k})^2 \right ] \delta (w - E_i({\bf k})).
\end{equation}
The term proportional to $U_{1i}({\bf k})^2$ ($U_{3i}({\bf k})^2$)
gives the contribution to the density of states of the
first (second) sublattice made by the $i^{\mbox{th}}$
band.  It is therefore possible to write the DOS as the sum of the two
sublattice densities of states
\begin{equation}
  \rho(\omega) = \rho_1(\omega) + \rho_2(\omega).
\end{equation}
We will evaluate both the DOS and the gap equation, (\ref{10}), in
the following section.

%%%
\section{A Discussion of Numerical Results}
\label{sec-DOSII}
\subsection{Specification of the Model}
The goal of this section is to discuss the density of states of our
model.  We will do this by examining the features of our quasiparticle
energy dispersions.  To begin with however, we must choose
a specific form for our Hamiltonian (equation (\ref{1})).
We take
\begin{equation}
  \label{4}
  \xi_i({\bf k}) = -2 \sigma_i [ \cos(k_x) + \cos(k_y)] - \mu_i, \: i = 1,2
\end{equation}
which comes from assuming that the electrons are described by tight binding
dispersions within the planes. The Brillouin zone is $-\pi < k_x, k_y \le \pi,
\, -\pi/d < k_z \le \pi/d$ where $d$ is the lattice constant in the
$z$--direction.  Since $\xi_i$ describes the energy of an electron in a
Bloch state of one of the sublattices, it will have a weak $k_z$ dependence.
However, a much larger $k_z$ dependence will come from $t({\bf k})$,
which describes the nearest neighbour hopping of electrons between
layers of different types.
The two chemical potentials in (\ref{4}) allow the two bands to be
offset from each other.  The most significant feature of this choice
of $\xi_i$ is that there are saddle points at $(k_x, k_y) = (0, \pi)$
(at which $\xi_i = -\mu_i$) and other symmetry related points.  A great
deal has been written about model systems in which the saddle points lie at
the Fermi surface where they play an important role in the dynamics of
the system \cite{saddlept}.  We wish to avoid this case since the aim
of this paper
is to understand the role of the interlayer coupling.  For all of the
results presented in this paper, then, we have taken $|\mu_1|, |\mu_2| >
\max(|t({\bf k})|, |\Delta_{\bf k}|)$.

The form of the interlattice coupling follows from the assumption that
that the electrons are tightly bound to the planes:
\begin{equation}
  \label{3}
  t({\bf k}) = t_1 e^{i k_z d_1} + t_2 e^{-i k_z d_2}.
\end{equation}
Here $t_1$ and $t_2$ are complex constants,
and the planes are alternately a distance $d_1$ and $d_2$ apart so that
$d = d_1 + d_2$.  The condition
$t({\bf k}) = t^\ast(-{\bf k})$ will be satisfied if $t_1$ and $t_2$ have
the same phase.  For our numerical calculations we reduce the
number of free parameters by setting $d_1 = d_2 = d/2$, $t_1 = t_2 = t_0/2$
so that
\begin{equation}
  \label{3a}
  t({\bf k}) = t_0 \cos (k_z d/2).
\end{equation}
One of the things which will become apparent is that, although the
fine structure of certain features of the DOS depend on the specific
form we choose for $t({\bf k})$, the existence and important properties
of these features are independent of this choice.

The pairing potential is taken to be separable so that
$V_{{\bf k k}^\prime} = V \eta_{\bf k} \eta_{{\bf k}^\prime}$,
with $ \eta_{\bf k} = 1$ for s--wave
superconductors and $ \eta_{\bf k} = \cos (k_x) - \cos (k_y)$ for
d--wave superconductors.  In this case we can define $\Delta_0$ by
\begin{equation}
  \label{3b}
  \Delta_0 = \frac{V}{\Omega} \sum_{\bf k} \eta_{\bf k} \langle
  {\bf a}_{1 -{\bf k} \downarrow} {\bf a}_{1 {\bf k} \uparrow}
  \rangle,
\end{equation}
and equation (\ref{10}) will simplify to
\begin{equation}
  \label{10a}
  \Delta_0 = \frac{V}{\Omega} \sum_{i=1}^4 \sum_{\bf k} \eta_{\bf k}
  U_{1i}({\bf k}) U_{2i}({\bf k}) f(-E_i({\bf k})),
\end{equation}
with $\Delta_{\bf k} = \Delta_0 \eta_{\bf k}$.  We solve equation (\ref{10a})
for $T_c$ and graph the results in Fig.\ \ref{Tc}.
We can see that the critical temperature decreases rapidly with
increasing $t_0$ which suggests that a weak coupling of the planes is
necessary for any reasonable description of a H$T_c$ superconductor.
It has been suggested \cite{Kresin,Bussmann}, however, that the inclusion
of phonon mediated interlayer coupling may actually lead to an increase
in $T_c$ with coupling strength.
We also calculate $\Delta_0(T=0)$ from equation (\ref{10a}) and plot
$\Delta_0(T=0)/T_c$ in Fig.\ \ref{Tc}.  We can see that $\Delta_0(T=0)/T_c$
increases rapidly for $t_0$ larger  than $.4\sigma_1$.
For YBCO the value of $2\Delta_{\mbox{max}}/T_c$ depends on the experiment
from which it is determined, but it is generally \cite{Gap} found to be
less than 8.
It is important to remember that the experimentally measured value of the gap
is typically the lowest energy peak in the quasiparticle excitation
spectrum, which is not
necessarily related in any simple fashion to the order parameter.
However, as a rough estimate, we can limit $\max(2\Delta_{\bf k})
= 4\Delta_0 < 8$ so that, from Fig.\ \ref{Tc}, $t_0/\sigma_1 < .5$.

Once $\Delta_0(T=0)$ is determined, we can proceed to calculate the
DOS for our Hamiltonian.  Our discussion of the DOS will be aided by
two observations.  The first is that equation (\ref{4}) implies that
\begin{equation}
  \label{10bb}
  \xi_2 = \frac{\sigma_2}{\sigma_1}\xi_1 + \frac{\mu_1\sigma_2-\mu_2\sigma_1}
  {\sigma_1}
\end{equation}
and the second is that the Brillouin zone boundary, $k_x,k_y = \pm\pi$ maps
into the diamond shaped boundary
\begin{equation}
  \label{10cc}
  |\Delta_{\bf k}| = 2\Delta_0 \left[ 1 - \frac{|\mu_i+\xi_i|}{4\sigma_i}
  \right]
\end{equation}
with $-4\sigma_i-\mu_i < \xi_i < 4\sigma_i-\mu_i$ and $i=1$ or $2$.
Equations (\ref{10bb}) and (\ref{10cc}) allow us to discuss the system
using $\xi_1$ (or $\xi_2$ for that matter) and $\Delta_{\bf k}$, rather
than $k_x$ and $k_y$, as independent variables.

We will consider two limiting cases of Eqn.\ (\ref{10bb}).  In the
first case $|\mu_1\sigma_2-\mu_2\sigma_1|/\sigma_1 \gg \Delta_0$ so that
$\xi_1^2+\xi_2^2 \gg \Delta_{\bf k}^2$ everywhere in the Brillouin zone.
In the second case, we take the Fermi surfaces of the sublattices to
coincide, so that
$\mu_1\sigma_2-\mu_2\sigma_1 = 0$.  The fundamental difference between
the two cases is that it is possible to distinguish band structure
effects from superconductivity effects in the first case, while in
the second case the interplane coupling plays an important role at
the Fermi surface.

\subsection{Case I:  Distinct Fermi Surfaces}
\label{sec-caseI}
We begin by discussing the quasiparticle energy dispersions.  First of all,
for most regions of the Brillouin zone $\xi_1^2 + \Delta_{\bf k}^2 -
\xi_2^2 \gg t^2, \Delta_{\bf k}^2$ and we see from Eqn.\
(\ref{7}) that $E_\pm$ approaches the limiting form
\begin{eqnarray}
\label{10b}
  E_+&\sim & \max(\sqrt{\xi_1^2 + \Delta_{\bf k}^2},|\xi_2 |) \nonumber \\
  E_-&\sim & \min(\sqrt{\xi_1^2 + \Delta_{\bf k}^2},|\xi_2 |).
\end{eqnarray}
In Fig.\ \ref{energy} we plot $E_\pm$ along the line $k_x = k_y$,
$k_z = 0$ and compare it with $|\xi_2|$ and $\sqrt{\xi_1^2 +
\Delta_{\bf k}^2}$.  We have chosen an s--wave gap for illustrative
purposes in the figure, since a d--wave gap would vanishe along this line.
In regions of the Brillouin zone where (\ref{10b}) is approximately true,
the effect of the matrix elements $U_{ij}$ will be to associate
$\sqrt{\xi_1^2 + \Delta_{\bf k}^2}$ with the first sublattice and
$|\xi_2|$ with the second sublattice, so that the integrands in equation
(\ref{11}) will become
\begin{eqnarray}
  &\frac{1}{2} \left [ 1 + \frac{\xi_1}{\sqrt{\xi_1^2 + \Delta_{\bf k}^2}}
  \right ] \delta(\omega - \sqrt{\xi_1^2 + \Delta_{\bf k}^2}) &
  \nonumber \\
  &+\frac{1}{2} \left [ 1 - \frac{\xi_1}{\sqrt{\xi_1^2 + \Delta_{\bf k}^2}}
  \right ] \delta(\omega + \sqrt{\xi_1^2 + \Delta_{\bf k}^2})&
\end{eqnarray}
for $\rho_1$ and
\begin{equation}
  \delta(\omega - \xi_2)
\end{equation}
for $\rho_2$.  For most energies, then, the DOS will simply look like the
DOS of the decoupled \cite{Zhou,Fedro} (small $t$) limit, which we show
in Fig.\ \ref{d1}

It is the regions where the interlayer coupling has an important effect
which we will discuss in the most detail.  In Figs.\ \ref{d3a}, \ref{d3}
and \ref{d2} we show the densities of states for nonzero interlayer
coupling.  These have a number of features that are absent in the
uncoupled case,
and we will discuss these features in turn.

The first thing which is clear is that, as $t_0$ is increased and the system
goes from being 2--dimensional to 3--dimensional, the logarithmic divergences
associated with the van Hove singularities (both the superconducting one
and those intrinsic to $\xi_1$ and $\xi_2$) become broadened peaks
\cite{Jelitto}.  This process is particularly clear in Figs.\ \ref{d3a} and
\ref{d3} where the intrinsic singularities at $\omega=.8$ in $\rho_1$ and
$\omega=-.4$ in $\rho_2$ are broadened considerably by $t_0$.

The interlayer coupling is also important in regions of
the Brillouin zone where $t^2$ is of the order of $\xi_1^2 +
\Delta_{\bf k}^2 - \xi_2^2$.  The effect of the coupling is to create
avoided band crossings wherever
\begin{equation}
  \label{10c}
  \xi_1^2 + \Delta_{\bf k}^2 - \xi_2^2 = 0.
\end{equation}
In Fig.\ \ref{energy},
for example, we can see that there are two avoided
crossings, one at $k_x=k_y=1.57$, and one at $k_x=k_y=1.32$.

At one of the avoided crossings ($k_x=k_y=1.57$ in Fig.\ \ref{energy}),
$\xi_1$ and $\xi_2$ have the same sign and (expanding $E_\pm$ in powers
of $\Delta_0$)
\begin{equation}
  E_\pm \sim |\xi^\prime| \pm |t({\bf k})| + O(\frac{\Delta_0^2}{\xi^\prime}),
\end{equation}
where $\xi_1 = \xi^\prime$ is the solution to Eqns.\ (\ref{10c}) and
(\ref{10bb}) for which $\xi_1$ and $\xi_2$ have the same sign.
The energy difference of the bands at the avoided crossing is
$\sim 2|t({\bf k})|$.  We can see from Fig.\ \ref{energy} that this
avoided crossing does not introduce
a van Hove singularity and, as a result, we do not find any feature in
the DOS of Figs.\ \ref{d3a} and \ref{d3} at this energy.
On the other hand, in Fig.\ \ref{d2}, we find that, because we have
let $\mu_1=\mu_2$, the avoided crossing occurs at the intrinsic van
Hove singularities of $\xi_1$ and $\xi_2$.  The result of this is that
the logarithmic peaks evident in Fig.\ \ref{d1} at $\omega \sim .8$ are
reduced far more than in Fig.\ \ref{d3}, where the avoided crossing
is far from the singularities.  This is an important result because it
shows that models in which the van Hove singularity plays an important
role (ie. it is near the Fermi surface) cannot also have a band
crossing near the Fermi surface.

For the second avoided band crossing (at $k_x=k_y=1.32$ in Fig.\
(\ref{energy})), $\xi_1$ and $\xi_2$ have opposite sign.  Again we
can expand $E_\pm$ in powers of $\Delta_0$, and we find that the
avoided crossing occurs at
\begin{equation}
  \label{11aa}
  E_\pm = \sqrt{\xi^{{\prime\prime}2} + t({\bf k})^2} \pm \frac{|t({\bf k})
  \Delta_{\bf k}|}
  {\sqrt{\xi^{{\prime\prime}2} + t({\bf k})^2}} +
O(\frac{\Delta_0^2}{\xi^{\prime\prime}})
\end{equation}
($\xi_1=\xi^{\prime\prime}$ is the solution to (\ref{10c}) and (\ref{10bb})
for which $\xi_1$ and $\xi_2$ have opposite sign).
We can see from Fig.\ \ref{energy} that there is actually a band gap of width
\begin{equation}
  \label{11a}
  E_+ - E_-\sim \frac{2|t({\bf k})\Delta_{\bf k}|}
  {\sqrt{\xi^{{\prime\prime}2} + t({\bf k})^2}}+
O(\frac{\Delta_0^2}{\xi^{\prime\prime}})
\end{equation}
at this point.  What is interesting
about this gap is that it only exists in the superconducting state
(since it is proportional to $\Delta_{\bf k}$) and that it forms {\em
away} from the Fermi surface.  This is unusual since the effects of
superconductivity normally manifest themselves near the Fermi surface.
In Fig.\ \ref{d3a}, these structures are clearly visible at $\omega
\sim \xi^{\prime\prime} = 0.55$ in the first sublattice and $\omega \sim
-\xi^{\prime\prime}$ in the second sublattice.  In Fig.\ \ref{d3}, the
larger value of $t_0$ shifts the location of the minima slightly to
$\omega = \pm 0.65$.  In Fig.\ \ref{d2}, the structures are
clearly visible at $\omega = \pm 0.4$.
The widths of the gap--like structures are given
approximately by
\begin{equation}
  \label{11b}
  \left. E_+-E_- \rule{0pt}{16pt}\right|_{\mbox{max}} \sim
  \frac{4t_0\Delta_0}{\sqrt{\xi^{{\prime\prime}2}+
  t_0^2}}\left[ 1 - \frac{|\mu_1+\xi^{\prime\prime}|}{4\sigma_1} \right]
\end{equation}
since the maximum value of $\Delta_{\bf k}$ allowed for
$\xi_1=\xi^{\prime\prime}$ is given by Eqn.\ (\ref{10cc}).
The source of this gap is not immediately obvious, but the proportionality
of Eqn.\ (\ref{11b}) to $t_0\Delta_0$ suggests that it results from an
interference of interlayer coupling and the mean field.  In section
\ref{sec-gap} we will show that this is in fact the case, and we will
discuss why these gaps appear on opposite sides of the Fermi surface
in each of the sublattices.

The final place where the interlayer coupling is important is near the
Fermi surface, where it influences the superconducting properties of
the system.
In Fig.\ \ref{energy} we see that $E_-$ has two local
minima which are related to the superconducting gaps
of the two sublattices.  We expand $E_-$ (Eqn.\ (\ref{8})) in
$(t^2-\xi_1\xi_2)$ and $\Delta_{\bf k}^2$ to get
\begin{equation}
  \label{12}
  E_-^2 \sim \frac{(t^2 - \xi_1 \xi_2)^2 + (\Delta_{\bf k} \xi_2)^2}
  {\xi_1^2 +  \xi_2^2 + 2 t^2}.
\end{equation}
Equation (\ref{12}) will vanish whenever $\Delta_{\bf k} = 0$ and
$t({\bf k})^2 = \xi_1\xi_2$.  This happens for two values of $\xi_i$,
which we denote by $\xi_i^{(j)}$ with $j=1,2$.
{}From Eqn.\ (\ref{10bb}) it is straightforward to show that
\begin{mathletters}
\label{12aa}
\begin{eqnarray}
  \xi_1^{(j)} & = & \frac{\mu_2 \sigma_1 - \mu_1 \sigma_2}{2\sigma_2}
  \nonumber \\
  &&+ (-1)^j \frac{\sqrt{(\mu_2 \sigma_1 - \mu_1 \sigma_2)^2 + 4 \sigma_1
  \sigma_2 t^2({\bf k})}}{2 \sigma_2} \\
  \xi_2^{(j)} & = & -\frac{\mu_2 \sigma_1 - \mu_1 \sigma_2}{2\sigma_1}
  \nonumber \\
  &&+ (-1)^j \frac{\sqrt{(\mu_2 \sigma_1 - \mu_1 \sigma_2)^2 + 4 \sigma_1
  \sigma_2 t^2({\bf k})}}{2 \sigma_1}.
\end{eqnarray}
\end{mathletters}
We then estimate the sizes of the superconducting gaps by substituting
the equations (\ref{10cc}), $\xi_i = \xi_i^{(j)}$ and $t = t_0$ into Eqn.\
(\ref{12}):
\begin{equation}
  \label{12bb}
  E_- \sim 2\Delta_0 \left| \frac{\xi_2^{(j)}}{\xi_1^{(j)}+\xi_2^{(j)}}
  \right| \left[1-\frac{|\mu_1+\xi_1^{(j)}|}{4\sigma_1} \right]
\end{equation}
In the limit that $t_0$ becomes small, the gap sizes become
\begin{eqnarray}
  \label{12cc}
  E_- &\sim &2\Delta_0\left[1-\frac{|\mu_1|}{4\sigma_1}\right] \nonumber \\
  && +\frac{\Delta_0 t({\bf k})^2}{2(\mu_2\sigma_1-\mu_1\sigma_2)} \left [
  \mbox{sgn}(\mu_1) - \frac{\sigma_1(4\sigma_1-|\mu_1|)}{\mu_2\sigma_1 -
  \mu_1\sigma_2} \right]
\end{eqnarray}
and
\begin{equation}
  \label{12dd}
  E_- \sim 2\Delta_0 \left [ \frac{\sigma_2 t_0}{\mu_2 \sigma_1 - \mu_1
  \sigma_2} \right ]^2 \left[1-\frac{|\mu_2|}{4\sigma_2}\right].
\end{equation}
Equation (\ref{12cc}) shows how a weak interlattice coupling affects
the gap in the intrinsically superconducting sublattice. One interesting point
is that whether $t_0$ increases or decreases the gap size depends on the
details of the band structure.  When $t_0=0$ we have an approximate
expression for the location of a logarithmic singularity in a two--dimensional
d--wave superconductor.  The exact expression \cite{Zhou} is
$\Delta_0(4\sigma_1-|\mu_1|)/ \sqrt{4\sigma_1^2+\Delta_0^2}$.  Equation
(\ref{12dd}) shows the size of the induced gap in the intrinsically normal
sublattice.

The gaps described by equations (\ref{12cc}) and (\ref{12dd}) will
appear in both $\rho_1$ and $\rho_2$ (this has also been pointed out
\cite{Bulaevskii} for a similar model).  The magnitudes of the
contributions
are determined by the matrix elements, $U_{ij}$, in Eqn.\ (\ref{11}).
In the appendix, we derive the low energy DOS and can see explicitely
how the matrix elements behave.  In particular, Eqns.\ (\ref{19a})
describe the strength with which each of the minima of $E_-$ (labelled
by $i=1,2$) is reflected in $\rho_1$ (from $U_1^2$) and $\rho_2$ (from
$U_2^2$).  In the weakly
coupled (small $t_0$) limit, for example, equations (\ref{12aa})
tell us that $U_1^2 \sim 1$, $U_2^2 \sim(\sigma_1 t)^2/
(\mu_2\sigma_1-\mu_1\sigma2)^2$ for the intrinsic gap given by (\ref{12cc}),
while $U_1^2 \sim(\sigma_2 t)^2
/(\mu_2\sigma_1-\mu_1\sigma2)^2$, $U_2^2 \sim 1$, for the induced gap given
by (\ref{12dd}).  In other words, features associated with one sublattice
are reflected in the other sublattice with a strength proportional to
$t_0^2$.  We can see quite clearly in Fig.\ \ref{d2}, for example,
that both gaps are reflected in both densities of states.

The main results of the appendix are expressions for the
low energy DOS in which a) two of the three k--space integrations have
been performed approximately (Eqn.\ \ref{20}) and b) the final
integration is performed in a crude approximation in the small $t_0$
limit to find $\rho_1$ (Eqns.\ (\ref{25urg})) and
$\rho_2$ (Eqns.\ (\ref{27urg})).  The important features of
these expressions can be summarised simply.  First of all, the quantity
$t^\ast (\omega)$ (Eqn.\ (\ref{24bb})) gives the energy scale of the
induced gap.  In fact, the
equation $t(k_z) = t^\ast (\omega)$ is just a rearrangement of the
expression (\ref{12dd}) for the induced gap.  Eqn.\ (\ref{25}) (with
$\lambda_\pm$ given by Eqns.\ (\ref{21}) and (\ref{22})) is
an approximate expression for the d--wave gap in the S sublattice.
Eqns.\ (\ref{25aa}) and (\ref{25bb}) are the corrections (of order
$t_0^2$) to $\rho_1$ inside and outside the induced gap respectively.
Similarly, Eqns.\ (\ref{27}) and (\ref{27aa}) give $\rho_2$ inside
and outside the induced gap respectively.  One of the most important
features of
the induced gap is that, since both $t(k_z)$ and $\Delta_{\bf k}$
have nodes, there are a large number of low energy excitations.  The
density of states still vanishes inside the induced gap but
instead of vanishing linearly, it vanishes with a divergent slope.
We can see this in Fig.\ \ref{d5} where we compare the low $\omega$
DOS, calculated numerically from Eqn.\ (\ref{11}), with the approximate
expressions derived in the appendix.

\subsection{Case II:  Coincident Fermi Surfaces}
\label{sec-caseII}
We will now consider the case where the Fermi surfaces of the two
sublattices coincide,  which requires that $\mu_1 \sigma_2 = \mu_2 \sigma_1$
so that Eqn.\ (\ref{10bb}) becomes
\begin{equation}
  \label{100}
  \xi_2 = \frac{\sigma_2 \xi_1}{\sigma_1}.
\end{equation}
Away from the Fermi surface, the structure of the DOS is that of the
uncoupled ($t_0=0$) limit for exactly the same reasons as in the
previous case.  Near the Fermi surface, however, the DOS behaves quite
differently and is somewhat more complicated to describe, though
a partially quantitative understanding is still possible.
In Fig.\ \ref{energy2} we can see that $E_-$ still has a
double minimum structure throughout much of the Brillouin zone but
that it is no longer possible to associate either of the minima with a
particular sublattice.  In Figs.\ \ref{d4} and \ref{d4a}, we show the DOS
for two different strengths of interlayer coupling.  In Fig.\ \ref{d4},
the density of states exhibits a nested gap structure at the Fermi
surface.  For larger values of $t_0$ (Fig.\ \ref{d4a}), the outer structure
is no longer identifiable as a gap and the DOS is similar in appearance
to that of a 2--dimensional d--wave material.
However, as we shall see, the dependence of
the gap width on the order parameter, $\Delta_0$, is vastly different
than in the 2--dimensional case.  The complicated structure of Figs.\
\ref{d4} and \ref{d4a} is due to the mixing of band structure and
superconductivity effects at the Fermi surface.  Because of this,
it is not possible to replicate the analysis of the last section where
we were able to identify gap--like structures in the DOS with features of the
energy spectrum.  Instead, it is necessary to resort to the brute--force
method of identifying saddle point singularities in $E_\pm$, and determining
which features in the figures they are responsible for.
The method used below is approximate and works well provided
that $\Delta_0/\sigma_1 \ll 1$.

Since finding the zeros of $\nabla_{\bf k}E_\pm$ is difficult, our
approach is to set $\partial E_\pm / \partial \xi_1 = 0$ to find the
surfaces of extrema
of $E_\pm$ with respect to $\xi_1$, and then to assume that the van
Hove singularities are at the extremal values of $|t({\bf k})|$
and $|\Delta_{\bf k}|$ on this surface.  The extremal values of $|t|$
are simply zero and $t_0$, which is actually what we would
find by setting $\nabla_{\bf k}E_\pm = 0$ anyway.  On the other hand,
the extremal values of $|\Delta_{\bf k}|$ are either zero, or given by
Eqn.\ (\ref{10cc}).  Since this last constraint depends on $\xi_1$,
it is wrong to treat $\Delta_{\bf k}$ as independent when taking
$\partial E_\pm / \partial \xi_1$.  Providing that $\Delta_0$ is
small, however, the approximation is good and introduces an error
of $O(\Delta_0^3/\sigma_1^2)$ to the energies of the van Hove singularities.

We begin by substituting equation (\ref{100})
into the expression for $E_\pm$ (equation (\ref{7})) and setting
$\partial E_\pm / \partial \xi_1 = 0$.
We find that there are two solutions:
\begin{equation}
  \label{101}
  \xi_1 = 0
\end{equation}
(for both $E_+$ and $E_-$), and
\widetext
\begin{equation}
  \label{102}
  \xi_1^2  = \frac{\sigma_1(\sigma_1^2 + \sigma_2^2) |t({\bf k})|
  \sqrt{t({\bf k})^2(\sigma_1+\sigma_2)^2 + 2 \sigma_2 \Delta_{\bf k}^2
  (\sigma_1 - \sigma_2)}
  - \sigma_1^2\sigma_2[\Delta_{\bf k}^2 (\sigma_1 - \sigma_2) +
  2 t({\bf k})^2 (\sigma_1 + \sigma_2)] }
  {\sigma_2(\sigma_1+\sigma_2)(\sigma_1-\sigma_2)^2}
\end{equation}
(for $E_-$ only). This second equation gives the location of the two
minima of $E_-$ and only has positive solutions for $\xi_1^2$ if
\begin{equation}
  \label{103}
  t({\bf k})^2 > \frac{\sigma_2 \Delta_{\bf k}^2 \left [ (\sigma_1 + \sigma_2)
  (\sigma_1^2
  +\sigma_2^2) - 2 \sigma_1^3 +  (\sigma_1^2+\sigma_2^2) \sqrt{\sigma_2^2 +
  2\sigma_1 \sigma_2 + 2\sigma_1^2} \, \right ]} {(\sigma_1 + \sigma_2)^4}.
\end{equation}
The important difference, then, between Figs.\ \ref{energy2}(a) and
\ref{energy2}(b) is the size of $|t({\bf k})|$.  This crossover in
behaviour was noted previously \cite{Buzdin}, for the slightly
simpler model in which $\xi_1\equiv\xi_2$ and $\Delta_{\bf k} \equiv
\Delta_0$.
Now at $\xi_1 = 0$,
\begin{equation}
  \label{104}
  E_\pm = \sqrt{\left ( \frac{\Delta_{\bf k}}{2} \right )^2 + t^2}
  \pm \left | \frac{\Delta_{\bf k}}{2} \right |,
\end{equation}
while when $\xi_1$ is given by (\ref{102}),
\begin{equation}
  \label{105}
  E_-^2 = \frac{2 |t({\bf k})| \sigma_1 \sigma_2
  \sqrt{t({\bf k})^2(\sigma_1+\sigma_2)^2 + 2 \sigma_2 \Delta_{\bf k}^2
  (\sigma_1 - \sigma_2)}
  - \sigma_2 [\Delta_{\bf k}^2 \sigma_2 (\sigma_1 - \sigma_2) +
  2 t({\bf k})^2 \sigma_1 (\sigma_1 + \sigma_2)] }
  {(\sigma_1+\sigma_2)(\sigma_1-\sigma_2)^2 } .
\end{equation}
\narrowtext

We first describe
the saddle points at which $\xi_1 = 0$.  The extremal values of $|t|$
are $|t|=0$ and $|t|=t_0$, and the extremal values of $\Delta_{\bf k}$
are zero and (from (\ref{10cc}))
\begin{equation}
  \label{106}
  \Delta^{\prime} = 2 \Delta_0 \left [ 1-\frac{|\mu_1|}{4\sigma_1} \right ] .
\end{equation}
Of the four points in the Brillouin zone just described, only two are
actually saddle points (as opposed to band minima).  There will be a
saddle point in $E_+$ at
$\Delta_{\bf k} = \Delta^\prime$, $t({\bf k}) = 0$ at which
\begin{equation}
  \label{107}
  E_+ = \Delta^\prime.
\end{equation}
There will also be saddle points in $E_+$ and $E_-$ at $\Delta_{\bf k} =
\Delta^\prime$, $t({\bf k}) = t_0$ at which
\begin{eqnarray}
  \label{108a}
  E_- & = & \sqrt{ \left( \frac{\Delta^\prime}{2} \right )^2 + t_0^2}
  - \frac{\Delta^\prime}{2} \\
  \label{108b}
  E_+ & = & \sqrt{ \left( \frac{\Delta^\prime}{2} \right )^2 + t_0^2}
  + \frac{\Delta^\prime}{2}.
\end{eqnarray}

The double minimum, with energy given by (\ref{105}) will also have
two saddle points, though they are somewhat more complicated to describe.
First of all, there will only be saddle points if the inequality (\ref{103})
is satisfied for $t = t_0$ and $\Delta = \Delta^{\prime\prime}$.
The quantity, $\Delta^{\prime\prime}$ is the solution to the pair of equations,
(\ref{102}) and (\ref{10cc}) with $t({\bf k}) = t_0$.  These
equations can be solved iteratively in a straightforward fashion, and
will yield {\em two} closely
spaced solutions corresponding to the two slightly different energies
at which the minima of $E_-$ intersect the Brillouin zone boundary.
The energy of the saddle point is then given by Eqn.\ (\ref{105}).

In Fig.\ \ref{d4}, $\Delta^\prime = 0.14$, and
the singularities at $\pm\Delta^\prime$ can be associated with the outer
gap in the nested gap structure.  If we recall Eqn.\ (\ref{12cc}) and
the discussion which follows, we will recognize that $\Delta^\prime$
is approximately the energy of the gap in a 2--dimensional d--wave
superconductor.  In Fig.\ \ref{d4a}, however, the nested gap structure
is no longer clear and, and although there are features at $\pm
\Delta^\prime = \pm 0.23$, they do not appear as a
gap--like structure.  In Fig.\ \ref{d4a}, the gap width is actually
given by Eqn.\ (\ref{105}) which gives the closely spaced sets of
singularities $\omega = \pm0.072$ and $\omega = \pm0.096$.  Eqn.\ (\ref{105})
makes it clear that the structure in the DOS which we naively associate
with the superconducting gap is, in fact, due to a complicated mixing
of the gap and the interlayer coupling.  It is
an important point that a tunneling experiment might not be able to
distinguish between Fig.\ \ref{d4a} and the density of states
of a 2--dimensional d--wave superconductor ($\rho_1$ in Fig.\ \ref{d1}).
In this case, then, a tunneling experiment would reveal very little
about either the interlayer coupling or the nature of the pairing
interaction.

In Fig.\ \ref{d4}, the singularities due to Eqn.\ (\ref{105}) are at
$\omega = \pm 0.0431$ and $\omega = \pm 0.0426$.  They are responsible
for the inner gap in the nested gap structure.  There is also a van
Hove singularity nearby, at $\omega = 0.05$, which comes from Eqn.\
(\ref{108a}).  It is difficult to see because of its close proximity
to the other singularities, but in Fig.\ \ref{d4a}, this feature is
clear and is at $\omega = \pm 0.30$.   Once again, this feature is the
result of the interference of interplane coupling with the mean field
and it cannot be considered a perturbation of either a superconducting
gap or a band gap.  In Fig.\ \ref{d4}, Eqn.\ (\ref{108b}) describes
the smearing of the logarithmic singularity due to the intrinsic
superconductivity of the first sublattice.  It gives van Hove singularities
at $\omega = \pm 0.195$.  This should be compared with the
second term of Eqn.\ (\ref{12cc}), whose $k_z$ dependence also smears
the logarithmic singularity associated with the intrinsic gap.  The
difference is that, in (\ref{12cc}), $t$ appears as a perturbation,
while in (\ref{108b}), the smearing is inherently non--perturbative.
In Fig.\ \ref{d4a}, the van Hove singularities are at $\omega =
\pm .534$.  The large broadening of the logarithmic singularity has
caused it to overlap adjacent features so that it is not obvious that
it is related to the singularity at $\Delta^\prime$.
The remaining features in the DOS of Figs.\ \ref{d4} and \ref{d4a}
are due to the inherent structure of $\xi_1$ and $\xi_2$.

\subsection{Discussion of the Superconducting Band Gap}
\label{sec-gap}
In section \ref{sec-caseI} we discussed the sublattice densities of states
in the limit $|\mu_1\sigma_2-\mu_2\sigma_1|\gg \Delta_0\sigma_1$.
We found that, in addition to the expected band and superconducting gaps,
there were gap--like structures which appeared in the superconducting state
{\em away} from the Fermi surface.  This is surprising since the effects
of superconductivity normally manifest themselves at the Fermi surface.
A further property of these structures is that, if in one sublattice a
gap--like structure should appear at some energy, $\omega$, then there
will be a similar structure in the other sublattice at energy $-\omega$.
The purpose of this section is to understand the source of these
unusual gap--like structures.

The first clue to the nature of the structures is in the expression
(\ref{11b}), which shows that the gap width is proportional to
$\Delta_0 t_0$.  This suggests that there is an interference between
the interlayer coupling and the mean field.  In order to examine
this interference further, we will write out an equation for the
Green's function in which the interlayer coupling and the mean field
are treated as perturbations.
The Hamiltonian, Eqn.\ (\ref{4a}), can be written as
\begin{equation}
  \label{110}
  {\bf H} - {\bf N}\mu = \sum_{\bf k} {\bf A}^\dagger ({\bf k})
  \{ Q^0 + Q^1 \}{\bf A}({\bf k}) + \mbox{const}
\end{equation}
where
\begin{equation}
  \label{110a}
  Q^0 =  \left [ \begin{array}{cccc}
  \xi_1({\bf k}) & 0 & 0 & 0 \\
  0 & -\xi_1(-{\bf k}) & 0 & 0 \\
  0 & 0 &  \xi_2({\bf k}) & 0 \\
  0 & 0 & 0 & -\xi_2(-{\bf k})
  \end{array} \right ]
\end{equation}
and
\begin{equation}
  \label{111}
  Q^1 =  \left [ \begin{array}{cccc}
  0 & -\Delta_{\bf k} & t({\bf k}) & 0 \\
  -\Delta^\ast_{\bf k} & 0 & 0 & -t^\ast(-{\bf k}) \\
  t^\ast({\bf k}) & 0 & 0 & 0 \\
  0 &  -t(-{\bf k}) & 0 & 0
  \end{array} \right ] .
\end{equation}
Equation (\ref{111}) is treated like a perturbation.  The temperature
Green's function is
defined as $G_{ij}({\bf k}; \tau) = - \langle T {\bf A}_i({\bf k},-i\tau),
{\bf A}^\dagger_j({\bf k}, 0) \rangle$, where $T$ is the time--ordered product
and the time dependence of ${\bf A}_i({\bf k},t)$ is determined by the
full Hamiltonian, $Q$. It is straightforward to show that
\begin{equation}
  \label{112}
  G_{ij}(i\zeta_n) = G_{ij}^0(i\zeta_n) + \sum_{l,m=1}^{4}
  G_{il}^0(i\zeta_n)Q^1_{lm}G_{mj}(i\zeta_n).
\end{equation}
where $G(i\zeta_n)$ and $G^0(i\zeta_n)$ are the Fourier components of
$G(\tau)$ and $G^0(\tau)$ and $\zeta_n = \pi(2n+1)/\beta$ are the
Matsubara frequencies.  The uncoupled Green's function, $G^0_{ij}$,
is the full Green's function with $t_0=\Delta_0=0$.  It is easy to
show that $G_{ij}^0(i\zeta_n) = \delta_{ij}[i\zeta_n - Q^0_{ii}]^{-1}$.

With a little work we are able to write out equation (\ref{112}) for
$G_{11}$ explicitely:
\begin{mathletters}
\begin{eqnarray}
  \label{113}
  G_{11} &=& G^0_{11} + G^0_{11}\Delta_{\bf k}G^0_{22}\Delta_{\bf k}^\ast
  G_{11} + G^0_{11}tG^0_{33}t^\ast G_{11} \nonumber \\
  & &+ G^0_{11}\Delta_{\bf k}G^0_{22}t^\ast G_{41}
\end{eqnarray}
\begin{equation}
  \label{114}
  G_{41} = G^0_{44}tG^0_{22}\Delta^\ast_{\bf k}G_{11} +
  G^0_{44}tG^0_{22}t^\ast G_{41}.
\end{equation}
\end{mathletters}
The Green's functions, $G_{11}$ and $G_{33}$, describes the propagation of
spin--up electrons in the first and second sublattices respectively.  On the
other hand, $G_{22}$ and $G_{44}$ describe
the propagation of spin--down holes in the first and second sublattices.
The unusual term, $G_{41}$, is essentially defined by Eqn.\ (\ref{114})
and it describes an electron which both hops between the planes and
interacts with the mean field.

In Figs.\ \ref{f1}(a) and \ref{f1}(b) we write Eqns.\ (\ref{113}) and
(\ref{114}) out in
diagrammatic form.  In part (a) we see that there are three processes
which modify the propagation of an electron.   The first of these
processes, which simply describes the hopping of electrons between adjacent
layers, would appear in any band structure calculation involving two atoms
per unit cell.  The effect of this term is to cause the two bands,
$\xi_1$ and $\xi_2$, to repel each other wherever they cross so that we
have avoided band crossings whenever $\xi_1=\xi_2$.  The second process
in Fig.\ \ref{f1}(a) describes the intrinsic superconductivity of
the first sublattice.  It is useful, however to interpret this in the same
way we interpreted the first process:  as a coupling of two bands.
The important
point here, however, is that the two bands which are coupled by the mean
field are not the two sublattice bands, but are, instead, the spin--up
and spin--down components of the band in the first sublattice.  Moreover,
the coupling is actually between spin-up electron states and
spin--down hole states.
The terms in the Hamiltonian responsible for the superconductivity can
be interpreted as describing transitions between the electron and hole
bands in exactly the same way that the terms responsible for interlayer
coupling describe transitions between electron bands.
This is clear in Fig.\ \ref{f1}(a) where an interaction with the mean
field changes an electron into a hole.
The second term in Fig.\ \ref{f1}(a), then, will cause an avoided
crossing in the quasiparticle energies wherever the original electron
and hole bands cross.  Since the dispersion of the original hole band is just
$-\xi_1$, the superconducting band gap will open up at $\xi_1=-\xi_1=0$.
We can see that the reason that the effects of superconductivity
normally manifest themselves at the Fermi surface is that the interaction
between the hole and electron bands is largest there.

The final term in Fig.\ \ref{f1}(a) is interesting because it describes
the interference of the mean field and interlayer coupling.
In (c) we have iterated this once, using Fig.\ \ref{f1}(b), to show the
lowest order contribution
to the mixing.  The diagram in (c) describes an electron of energy $\xi_1$
which interacts with the mean field to become a hole of energy $-\xi_1$
and then hops to the second sublattice before reversing the process.
The effect of this diagram is simply to couple electron states in the
first sublattice with hole states in the second sublattice.  Exactly
as before then, we should expect avoided crossings in our quasiparticle
energy dispersions whenever $\xi_1 = -\xi_2$.  As we discussed in
section \ref{sec-caseI}, this avoided crossing is responsible for the
gap--like structures which appear away from the Fermi surface in Figs.\
\ref{d3a}, \ref{d3} and \ref{d2}.

In section \ref{sec-caseII} we discussed the case of coincident sublattice
Fermi surfaces.  In this case, Fig.\ \ref{f1}(a)
still describes the system, but all three processes will contribute to the
avoided band crossings at the same time.  In particular, the final
process in (a) will be responsible for the complicated nature of
the gap at the Fermi surface.

%%%
\section{Conclusion}
\label{sec-conc}

In this paper we have presented results on the density of states of
a layered S/N system.  The inherently superconducting sublattice
was presumed to have a gap with d--wave structure.  We investigated
how the (coherent) coupling between the two types of
layers changed the DOS from the usual 2--dimensional case.
Our chosen Hamiltonian, Eqn.\ (\ref{1}), led to a relatively complicated
quasiparticle energy dispersion (Eqn.\ (\ref{7}) or (\ref{8})), which
we examined in an attempt to understand our numerical
results for the DOS.  We
found that the behaviour of our model could be summarised by two limiting
cases, distinguished by the relative positions of the sublattice Fermi
surfaces.

In the first case, the Fermi surfaces of the S and N sublattices were far
enough apart in the Brillouin zone that the interlayer coupling did not
cause them to interfere with each other.  In this limit it was clear that
the effect of the interlayer coupling was to shift the usual 2--dimensional
d--wave gap in the S sublattice perturbatively (Eqn.\ (\ref{12cc})) and
to induce a small gap at the Fermi surface of the N sublattice (Eqn.\
(\ref{12dd})).  The gaps in each of the sublattices were also weakly reflected
in the DOS of the other sublattice, with a magnitude proportional to
$t_0^2$.  In the appendix, we derived an approximate
expression for the density of states within the induced gap
(Eqn.\ \ref{27}), and we found that the behaviour is much different
from the linear behaviour found in the unperturbed d--wave case.
The most interesting features, though, of the DOS in this case were
the gap--like structures which appeared away from the Fermi surface
in the superconducting state (this was clearly visible in Fig.\ \ref{d3a}
and we gave an expression for the gap widths in Eqn.\ (\ref{11a})).
In section \ref{sec-gap}, we discussed this matter in some detail.
Essentially, the point was that there is an interference of the
interplane coupling and the mean field.  Since the mean field can
be viewed as an interaction between holes and electrons within the
superconducting sublattice, and the interplane coupling is an interaction
between electrons in the two sublattices, the interference of the
terms results in an interaction of holes and electrons in different
sublattices.  It is this interaction which resulted in the gap--like
structures.

In the second case we considered, the sublattice bands ($\xi_1$ and
$\xi_2$) crossed at their Fermi surfaces.  In this case, there was a
nontrivial mixing of mean field and superconductivity effects at the
Fermi surface.  In Fig.\ \ref{d4}, we showed that, for weak interlayer
coupling, the system has a nested gap structure, where the outer
gap is associated with the usual d--wave gap, and the innner gap
is a complicated function of the coupling and order parameter
(Eqn.\ (\ref{105})).  As the coupling was increased, however (Fig.\
\ref{d4a}), the outer gap became unrecognizable and the total DOS
became very similar to that for a single layer system in which there
is only one gap.

Perhaps the most important generalization which follows from this paper,
then, is that the interference of the interlayer coupling and the
superconducting mean field can
produce surprising behaviour.  One implication of this is that for a system
like YBa$_2$Cu$_3$O$_7$, where gap--like features have been observed
in tunneling and Raman scattering experiments, it could be very
wrong to simply associate these features directly with the order
parameter.

%%%
\section*{Acknowledgements}
This work was supported in part by the Natural Sciences and Engineering
Council of Canada (NSERC) and the Canadian Institute for Advanced Research
(CIAR).  W.A.A. was supported by an Ontario Graduate Scholarship (OGS).

%%%
\appendix
\section*{Low Energy DOS}
In this section we will derive the low energy approximation for
the density of states in the case in which $|\sigma_1\mu_2-\sigma_2\mu_1|/
(\sigma_1+\sigma_2) \gg \Delta_{\bf k},\, \omega$.  One immediate
consequence of this condition is that we can ignore terms in the DOS
(equation (\ref{11})) containing $\delta (w \pm E_+)$ since
\begin{equation}
  E_+^2 \ge \max(\xi_1^2,\xi_2^2) \ge \left [
  \frac{\sigma_1\mu_2-\sigma_2\mu_1}{\sigma_1 + \sigma_2} \right ]^2.
\end{equation}
This leaves us to consider the contributions to the DOS made by $E_-$.
Eqn.~(\ref{12}) is our first approximation to $E_-$ at low energies and we
recall
that it is zero whenever $\xi_1=\xi_1^{(i)},\, \xi_2=\xi_2^{(i)}$, and
$\Delta_{\bf k}=0$ with $i=1,2$ and $\xi_j^{(i)}$ given by Eqn. (\ref{12aa}).
We will proceed by considering the contributions from the neighbourhoods
of the two zeros of $E_-$ separately, and we break
the integral in Eqn. (\ref{11}) into a sum
of two integrals which are centred about each of these zeros.
We define the coordinates
\begin{equation}
  \label{14}
  \eta^{(i)} =  \xi_1 - \xi_1^{(i)}
\end{equation}
and expand $E_-$ to lowest order in $\eta^{(i)}$ and $\Delta_{\bf k}$:
\begin{equation}
  \label{15}
  E_-^2 \sim a^{(i)2} \eta^{(i)2} + b^{(i)2} \Delta_{\bf k}^2
\end{equation}
  with
\begin{eqnarray}
  a^{(i)} & = & \frac {\sigma_1  \xi_2^{(i)} + \sigma_2  \xi_1^{(i)}}
  {\sigma_1 [ \xi_1^{(i)} + \xi_2^{(i)}]} \nonumber \\
  b^{(i)} & = & \frac {\xi_2^{(i)2} }{t^2+ \xi_2^{(i)2}}. \nonumber
\end{eqnarray}

The Jacobian for the transformation $(k_x,k_y) \rightarrow (\eta^{(i)},
\Delta)$ (where $\Delta \equiv \Delta_{\bf k}$) is
\begin{eqnarray}
  \label{16}
  J & = & \frac{1}{4 \sigma_1 \Delta_0 \sin (k_x) \sin (k_y)} \nonumber \\
    & = & \frac{\Delta_0}{\sigma_1 \sqrt{[\alpha_+^2 - \Delta^2]
[\alpha_-^2 - \Delta^2]}}
\end{eqnarray}
with
\[
  \alpha_\pm = 2 \Delta_0 \left [ 1 \pm \frac{| \eta^{(i)}+\mu_1+\xi_1^{(i)}|}
  {4 \sigma_1} \right ].
\]
This transformation maps the square $0 < k_k, k_y \le \pi$ into a diamond
shaped region whose boundaries are given by the lines $\Delta =
\pm \alpha_-$, $-4\sigma_1-\mu_1-\xi_1^{(i)} \le \eta^{(i)} \le
4\sigma_1-\mu_1-\xi_1^{(i)}$.  Then
\begin{equation}
  \label{17a}
  \int_0^\pi dk_x  \int_0^\pi dk_y \rightarrow
  \int_{-4\sigma_1-\mu_1-\xi_1^{(i)}}^{4\sigma_1-\mu_1-\xi_1^{(i)}}
  d\eta^{(i)} \int_{-\alpha_-}^{\alpha_-} d\Delta \, J(\eta^{(i)},\Delta).
\end{equation}
Since we are interested in small $\omega$, we make the further approximation
that
\begin{equation}
  \label{17}
  \alpha_\pm \sim 2 \Delta_0 \left [ 1 \pm \frac{|\mu_1+\xi_1^{(i)}|}
  {4 \sigma_1} \right ].
\end{equation}
\widetext
We can now interchange the order of integration in (\ref{17a}) so that
for small positive $\omega$,
\begin{eqnarray}
  \label{18}
  \rho_{j}(\omega) & = & \frac{16}{N} \sum_{k_x,k_y = 0}^{\pi}
  \sum_{k_z = 0}^{\pi/d} U_{(2j-1) 2}^2  \delta
  (w - E_-) \nonumber \\
  & \sim & \frac{2 \Delta_0 d}{\sigma_1 \pi^3} \sum_{i=1}^{2} \int_0^{\pi/d}
  dk_z \, \int_{-\alpha_-}^{\alpha_-} d\Delta \int_{-4\sigma_1-\mu_1}
  ^{4\sigma_1-\mu_1}
  d\eta^{(i)} \frac{ U_{(2j-1) 2}^2  \delta (w - E_-)}
  {\sqrt{[\alpha_+^2 - \Delta^2][\alpha_-^2 - \Delta^2]}}.\nonumber \\
\end{eqnarray}
\narrowtext
Finally, in the limit of small $\omega$,
\begin{mathletters}
\label{19}
\begin{eqnarray}
   U_{12}^2 & \sim & \frac{1}{2} \frac{\xi_2^{(i)2}}{t^2 + \xi_2^{(i)2}}
  \left [ 1 - \frac{a^{(i)} \eta^{(i)}}{\omega} \right ] \\
   U_{32}^2 & \sim & \frac{1}{2} \frac{t^2}{t^2 + \xi_2^{(i)2}}
  \left [ 1 + \frac{a^{(i)} \eta^{(i)}}{\omega} \right ].
\end{eqnarray}
\end{mathletters}
The term $a^{(i)} \eta^{(i)} / \omega$ is not small but it is odd in
$\eta^{(i)}$ and will vanish in the integration.

We perform the integral over  $\eta^{(i)}$ in (\ref{18}) to get
\widetext
\[
  \rho_{j}(\omega) = \frac{2 d \Delta_0 \omega}{\sigma_1 \pi^3} \sum_{i=1}^{2}
  \int_0^{\pi/d} dk_z \, \frac{U_{j}^2}
  {|a^{(i)}b^{(i)}|} \int_{-\Delta_{\max}}^{\Delta_{\max}} \frac{d\Delta}
  {\sqrt{[\alpha_+^2 - \Delta^2][\alpha_-^2 - \Delta^2]
  [(\frac{\omega}{b^{(i)}})^2 - \Delta^2]}}
\]
\narrowtext
where $\Delta_{\max} = \min(\alpha_-,\omega/b^{(i)})$ and
\begin{mathletters}
\label{19a}
\begin{eqnarray}
  U_{1}^2 &=& \frac{\xi_2^{(i)2}}{t^2 + \xi_2^{(i)2}} \\
  U_{2}^2 &=& \frac{t^2}{t^2 + \xi_2^{(i)2}}.
\end{eqnarray}
\end{mathletters}
If $\Delta_{\max} = \omega/b^{(i)}$ then the energy surface $E_-=\omega$
is completely contained within the Brillouin zone.  The integral over
$\Delta$ can also be done analytically to give
\begin{equation}
  \label{20}
  \rho_{j}(\omega) = \frac{4 d \Delta_0 \omega}{\sigma_1 \pi^3} \sum_{i=1}^{2}
  \int_0^{\pi/d} dk_z \, \frac{U_{j}^2}{|a^{(i)}|}
  \frac{1}{\sqrt{\lambda_+}} K\left [ \frac{\lambda_+ - \lambda_-}
  {\lambda_+}\right].
\end{equation}
The function, $K(x)$, is the complete elliptic integral \cite{AbSt}.  If
$\omega < \alpha_+b^{(i)}$ then
\begin{mathletters}
\begin{eqnarray}
  \label{21}
  \lambda_\pm &=& \alpha_+^2 [ (\alpha_-b^{(i)})^2 + \omega^2]
  - 2 (\omega \alpha_-)^2 \nonumber \\
  && \pm 2\omega \alpha_-
  \sqrt{[(\alpha_+b^{(i)})^2 - \omega^2][\alpha_+^2 - \alpha_-^2]}
\end{eqnarray}
while if $\omega > \alpha_+b^{(i)}$ then
\begin{eqnarray}
  \label{22}
  \lambda_\pm &=& \omega^2[ \alpha_-^2 + \alpha_+^2]
  - 2 (\alpha_+\alpha_-b^{(i)})^2 \nonumber \\
  && \pm 2(\alpha_+\alpha_-) \sqrt{[\omega^2 - (\alpha_+b^{(i)})^2]
  [\omega^2 - (\alpha_-b^{(i)})^2]}. \nonumber \\
\end{eqnarray}
\end{mathletters}
There is nothing special about $\omega=\alpha_+b^{(i)}$ and the
integrand in (\ref{20}) is featureless at these points.  On the other
hand, $\lambda_-$ vanishes at $\omega=\alpha_-b^{(i)}$ and $K[
(\lambda_+-\lambda_-)/\lambda_+]$ diverges logarithmically.  The
integration over $k_z$ reduces the divergence to a peaked structure.
The magnitude of these peaks in each of the densities of states is
is dictated by $U_j^2$.

The complexity of (\ref{20}) prevents us from performing the final
integral analytically without further assumptions.  So far we have
used the fact that $\Delta_0$ and $\omega$ are small and we now add
to this the assumption that $t_0$ is small.  For one value of $i$,
\begin{equation}
  \label{23}
  (-1)^i \mbox{sgn}(\mu_2\sigma_1 - \mu_1\sigma_2) =1,
\end{equation}
and
$\xi_1^{(i)} \sim (\mu_2\sigma_1 - \mu_1\sigma_2)/\sigma_2$,
$\xi_2^{(i)} \sim \sigma_2 t^2/(\mu_2\sigma_1 - \mu_1\sigma_2)$,
$a^{(i)}\sim \sigma_2/\sigma_1$, $b^{(i)}=U_1^2 \sim(\sigma_2 t)^2
/(\mu_2\sigma_1-\mu_1\sigma_2)^2$, $U_2^2 \sim 1$, and
$\alpha_\pm~\sim 2\Delta_0[1~\pm~|\mu_2|/4\sigma_2]$. For the remaining
value of $i$,
\begin{equation}
  \label{24}
  (-1)^i \mbox{sgn}(\mu_2\sigma_1 - \mu_1\sigma_2) = -1,
\end{equation}
and
$\xi_1^{(i)} \sim -\sigma_1 t^2/(\mu_2\sigma_1 - \mu_1\sigma_2)$,
$\xi_2^{(i)} \sim -(\mu_2\sigma_1 - \mu_1\sigma_2)/\sigma_1$,
$a^{(i)}\sim b^{(i)}=U_1^2\sim 1$, $U_2^2 \sim(\sigma_1 t)^2
/(\mu_2\sigma_1-\mu_1\sigma_2)^2$, and $\alpha_\pm \sim
2\Delta_0[1\pm|\mu_1|/4\sigma_1]$.  Substituting this into (\ref{20})
gives
\begin{eqnarray}
  \label{24aa}
  \rho_1 &\sim& \frac{4\Delta_0\omega}{\sigma_1\pi^2}\frac{1}
  {\sqrt{\lambda_+}}K\left[ \frac{\lambda_+-\lambda_-}{\lambda_+} \right]
  \nonumber \\
  &+&\frac{4d\Delta_0\omega}{\pi^3}\frac{\sigma_2}{[\mu_2\sigma_1-
  \mu_1\sigma_2]^2}\int_0^{\pi/d} dk_z \, \frac{t(k_z)^2}{\sqrt{
  \lambda_+}}K\left[ \frac{\lambda_+-\lambda_-}{\lambda_+} \right].
  \nonumber \\
\end{eqnarray}
The first term is for $i$ given by (\ref{24}) and has $\lambda_\pm$
given by (\ref{21}) since $b^{(i)}\sim 1$.  The second term is for $i$
given by (\ref{23}) and, since $b^{(i)}$ is a function of $k_z$,
$\lambda_\pm$ will be given by (\ref{21}) when $t(k_z)>t^\ast$ and
by (\ref{22}) when $t(k_z)<t^\ast$.  The value of
$t^{\ast}$ comes directly from the condition $\omega=\alpha_+b^{(i)}$:
\begin{equation}
  \label{24bb}
  t^{\ast}(\omega) = \sqrt{\frac{2\omega}{\Delta_0\sigma_2
  [4\sigma_2 + |\mu_2|]}}\,|\mu_2\sigma_1-\mu_1\sigma_2|.
\end{equation}
In order to solve the final integral in (\ref{24aa}), we make the crude
approximation that
$\lambda_\pm$ can be replaced by the limiting forms
$\lambda_\pm(t\ll t^{\ast}) = \omega^2[\alpha_+ \pm \alpha_-]^2$ when
$t<t^{\ast}$, and $\lambda_\pm(t\gg t^{\ast}) = \alpha_+\alpha_-b^{(i)}$
when $t>t^{\ast}$.  Essentially, the point of this approximation is that
in regions of the Brillouin zone where the induced gap is smaller than
$\omega$, the second sublattice is treated as normal, while in the
remaining regions, $\omega$ is treated as much smaller than the induced
gap.  We find that $\rho_1(\omega) \sim \rho_1^\prime(\omega) +
\rho_1^{\prime\prime}(\omega)$ with
\begin{mathletters}
\label{25urg}
\begin{equation}
  \label{25}
  \rho_{1}^\prime(\omega) = \frac{4 \Delta_0 \omega}{\sigma_1 d \pi^2}
  \frac{1}{\sqrt{\lambda_+}}
  K\left [ \frac{\lambda_+ - \lambda_-}{\lambda_+}\right ]
\end{equation}
and
\begin{eqnarray}
  \label{25aa}
  \rho_1^{\prime\prime}(\omega)&= &
  \frac{\sigma_2t_0^2}{\pi^3[\mu_2\sigma_1-
  \mu_1\sigma_2]^2}K\left[1-\frac{\mu_2^2}{16\sigma_2^2}\right] \nonumber \\
  &\times & \left[\frac{\pi}{2} - \cos^{-1}(\frac{t^{\ast}}{t_0}) -
  \frac{t^{\ast}}{t_0}
  \sqrt{1-(\frac{t^{\ast}}{t_0})^2} \right] \nonumber \\
  &+&\frac{4 \sigma_2 t^{\ast2}}{\pi^2[\mu_2\sigma_1-\mu_1\sigma_2]^2}
  \left[1-\frac{|\mu_2|}{4\sigma_2}\right]^{-1}
  \cos^{-1}(\frac{t^{\ast}}{t_0})
\end{eqnarray}
when $t^\ast(\omega)<t_0$, and
\begin{equation}
  \label{25bb}
  \rho_1^{\prime\prime}  = \frac{\sigma_2t_0^2}{2\pi^2[\mu_2\sigma_1-
  \mu_1\sigma_2]^2}K\left[1-\frac{\mu_2^2}{16\sigma_2^2}\right]
\end{equation}
\end{mathletters}
when $t^\ast(\omega)<t_0$.
Equation (\ref{25}) is just the DOS which would result in the decoupled
($t_0=0$) limit.  Equation (\ref{25aa}) shows that the interlayer coupling
induces an inner gap for $t^\ast(\omega)<t_0$ (which is the prediction
made in Eqn. (\ref{12dd})).  The width of this gap is proportional
to $t_0^2$ and is difficult to see for weak interlattice coupling.

We can find the DOS in the second sublattice, $\rho_2$, in an analogous
manner:
\begin{mathletters}
\label{27urg}
\begin{eqnarray}
  \label{27}
  \rho_2(\omega)&\sim& \frac{2}{\sigma_2\pi^3}K\left[1-\frac{\mu_2^2}
  {16\sigma_2^2}\right]\left[\frac{\pi}{2} - \cos^{-1}(\frac{t^{\ast}}{t_0})
  \right] \nonumber \\
  &+& \frac{4 t^{\ast2}}{\sigma_2\pi^2t_0^2}
  \left[ 1- \frac{|\mu_2|}{4\sigma_2}\right]^{-1} \sqrt{
  (\frac{t_0}{t^\ast})^2-1} \nonumber \\
  &+& \frac{\omega t_0^2\sigma_1}{4\pi\Delta_0}[\mu_2\sigma_1-\mu_1
  \sigma_2]^{-2}\left[ 1- \frac{\mu_1^2}{16\sigma_1^2}\right]^{-1}
\end{eqnarray}
when $t^{\ast} < t_0$ and
\begin{eqnarray}
  \label{27aa}
  \rho_2(\omega)&\sim& \frac{1}{\sigma_2\pi^2}K\left[1-\frac{\mu_2^2}
  {16\sigma_2^2}\right] \nonumber \\
  &+& \frac{\omega t_0^2\sigma_1}{4\pi\Delta_0}[\mu_2\sigma_1-\mu_1
  \sigma_2]^{-2}\left[ 1- \frac{\mu_1^2}{16\sigma_1^2}\right]^{-1}
\end{eqnarray}
\end{mathletters}
when $t^{\ast} > t_0$.
Again we see that there is an induced gap for $t^\ast(\omega)<t_0$,
due to the interactions
of the sublattices.  For $t^\ast(\omega)>t_0$, $\rho_2(\omega)$ is just
the normal state DOS at the Fermi surface with corrections of order
$t_0^2$ due to the adjacent layers.
In Fig.  \ref{d5} we graph the low energy DOS calculated from equations
(\ref{11}), (\ref{20}) and from our analytical approximations.
We see that, although our analytic expressions for $\rho_1$ and
$\rho_2$ tend to overestimate the size of the inner gap, they
contain the essential description of the DOS.

%%%

%%%
\begin{figure}
\caption{$T_c$ and $\Delta_0(T=0)/T_c$ vs $t_0$.  The model parameters are
$\sigma_1=1$, $\sigma_2=0.6$, $\mu_1=-0.8$, $\mu_2=0.4$}
\label{Tc}
\end{figure}

\begin{figure}
\caption{Quasiparticle energy dispersion along the line $k_x=k_y$, $k_z=0$.
We have chosen an s--wave gap for purposes of illustration since a d--wave
gap vanishes along this line.
The features which are important for our discussion
of the density of states are the two minima of $E_-$, which are related
to the intrinsic and induced gaps, and the two avoided band crossings.
The parameters are $\sigma_1=1$, $\sigma_2
=0.6$, $\mu_1 = \mu_2 = -0.8$, $t_0 = 0.2$, $\Delta_0 = 0.1$}
\label{energy}
\end{figure}
\begin{figure}
\caption{Density of states in the normal and superconducting sublattices
for
the uncoupled limit.  There are logarithmic divergences at $\omega \sim
-\mu_1,-\mu_2$ due to the intrinsic van Hove singularities of $\xi_1$
and $\xi_2$.  There is also a d--wave gap at the Fermi surface in the
first sublattice.  The choices of parameters are $\sigma_1=1, \sigma_2
=0.6, \mu_1 = \mu_2 = 0.8, t_0 = 0.01,\Delta_0 = 0.1, T_c = 0.086$}
\label{d1}
\end{figure}
\begin{figure}
\caption{Density of states in the normal and superconducting sublattices.
Here the bands $\xi_1$ and $\xi_2$ cross away from their van Hove
singularities.  In addition to the intrinsic singularities at $\omega \sim
-\mu_1,-\mu_2$ and the d--wave gap in $\rho_1$ there are two gap--like
structures at $\omega \sim \pm 0.55$.  There is also an induced gap at the
Fermi surface in
$\rho_2$ which is too small to show on this plot. The parameters
are $\sigma_1=1$, $\sigma_2=0.6$, $\mu_1 = -0.8$, $\mu_2 = 0.4$,
$t_0 = 0.1$, $\Delta_0 = 0.09$, $T_c = 0.086$}
\label{d3a}
\end{figure}
\begin{figure}
\caption{Density of states in the normal and superconducting sublattices.
This figure shows the effect of increasing the interlayer coupling on
Fig.\ \protect\ref{d3a}.  The smearing of the logarithmic singularities
by the third dimension is clearly evident.  The induced gap at the Fermi
surface is now visible (although the fact that $\rho_2(0)=0$ is not clear
in the plot), and the gap--like structures away from the
Fermi surface are much larger, and shifted slightly, as we would expect
from Eqn.\ (\protect\ref{11aa}).  Here we have
$\sigma_1=1$, $\sigma_2=0.6$, $\mu_1 = -0.8$, $\mu_2 = 0.4$,
$t_0 = 0.4$, $\Delta_0 = 0.15$, $T_c = 0.086$}
\label{d3}
\end{figure}
\begin{figure}
\caption{Density of states in the normal and superconducting
sublattices.  In this case, the van Hove
singularities intrinsic to the sublattice dispersions at $\omega\sim
0.8$ are strongly suppressed by an avoided band crossing.  This should
be compared with Fig.\ \protect \ref{d3}, where the interlayer
coupling is the same, but the intrinsic van Hove
singularities of $\xi_1$ and $\xi_2$ can still be identified.
The other interesting feature of this figure is that we can see clearly
that the induced gap is reflected in $\rho_1$.  We should also note
again that $\rho_2$ actually vanishes at $\omega=0$.
Here $\sigma_1=1$, $\sigma_2=0.6$, $\mu_1 = \mu_2 = -0.8$,
$t_0 = 0.4$, $\Delta_0 = 0.15$, $T_c = 0.086$.}
\label{d2}
\end{figure}
\begin{figure}
\caption{Low frequency densities of states:  (a)  Determined numerically
from the exact expression, Eqn.\ (\protect \ref{11}) (b) Determined
numerically from the approximate expression, Eqn.\
(\protect \ref{20}) (c) Determined from the approximate analytical
expressions (\protect \ref{25urg}) and (\protect \ref{27urg}).  The model
parameters are $\sigma_1=1$, $\sigma_2=0.6$, $\mu_1=\mu_2=-0.8$, $t_0=0.2$,
$\Delta_0 = 0.15$.}
\label{d5}
\end{figure}
\begin{figure}
\caption{Quasiparticle energy dispersion along the lines (a)
$k_x=k_y$, $k_z=0$ and (b) $k_x=k_y$, $k_z=\pi$ for an s--wave gap.
In this case, the Fermi surfaces of $\xi_1$ and $\xi_2$ coincide.
Whether or not $E_-$, has a single minimum or double
minimum structure depends on the value of $t({\bf k})$ (see eqn.\
(\protect\ref{103})).
The parameters are $\sigma_1=1$, $\sigma_2=0.6$, $\mu_1 = -0.8$,
$\mu_2 = -0.48$, $t_0 = 0.2$, $\Delta_0 = 0.1$.}
\label{energy2}
\end{figure}
\begin{figure}
\caption{Density of states.
The Fermi surfaces of $\xi_1$ and $\xi_2$ are coincident.  In addition
to the intrinsic singularities at $\omega \sim -\mu_1,-\mu_2$, there
is a nested gap structure at the Fermi surface.  The outer gap is
related to the usual d--wave gap in a 2--dimensional superconductor
while the inner gap comes from a mixture of band structure and
superconductivity effects.
The parameters are $\sigma_1=1$, $\sigma_2=0.6$, $\mu_1 = -0.8$,
$\mu_2 = -0.48$, $t_0 = 0.1$, $\Delta_0 = 0.09$, $T_c = 0.086$.}
\label{d4}
\end{figure}
\begin{figure}
\caption{Density of states.  Again the
Fermi surfaces of $\xi_1$ and $\xi_2$ are coincident.  The increased
interlayer coupling has changed the appearance of the DOS considerably
from Fig.\ \protect\ref{d4}.  The outer gap no longer exists and
the gap which remains has a complicated dependency on both $t_0$ and
$\Delta_0$.  This DOS would not be simple to distinguish from that of
a 2--dimensional d--wave superconductor in a tunneling experiment,
even though the physics of the two systems is different. The
parameters are $\sigma_1=1$, $\sigma_2=0.6$, $\mu_1 = -0.8$,
$\mu_2 = -0.48$, $t_0 = 0.4$, $\Delta_0 = 0.15$, $T_c = 0.086$.}
\label{d4a}
\end{figure}
\begin{figure}
\caption{Feynman diagram representation of the equations
for $G_{11}$:  (a) and (b) are diagrammatic forms of Eqns.\
(\protect \ref{113}) and (\protect \ref{114}) respectively.
These equations form a self consistent set.  In (c), the final term
in (a) is iterated once, using (b), to give the lowest order effect
of the mixing of interlayer coupling and the mean--field.
Thick lines represent full Green's functions, thin lines represent
uncoupled Green's functions, rightward pointing arrows represent
spin up electrons, leftward pointing arrows represent spin down
holes and vertical dashed lines represent interlayer hopping.}
\label{f1}
\end{figure}

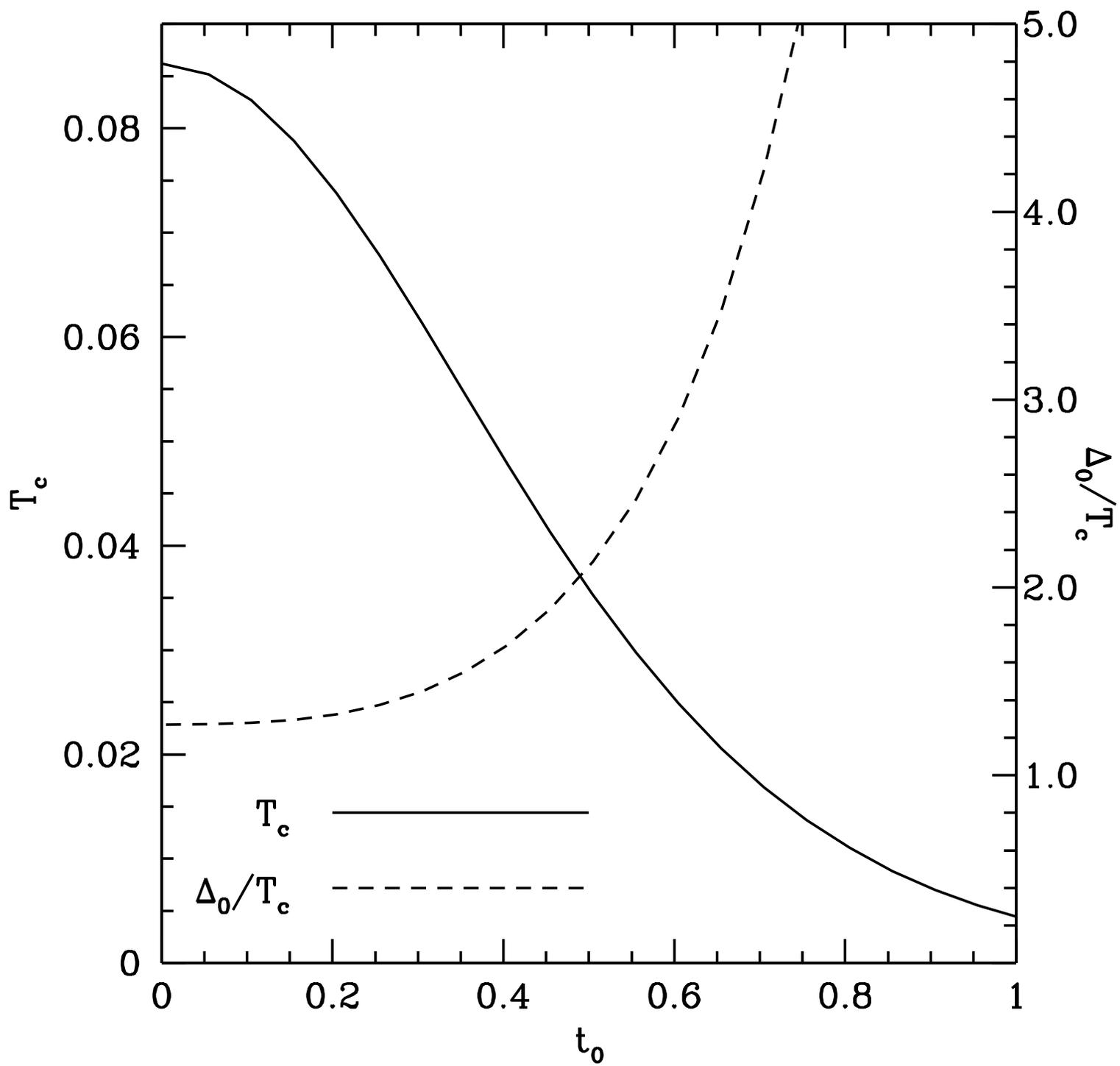
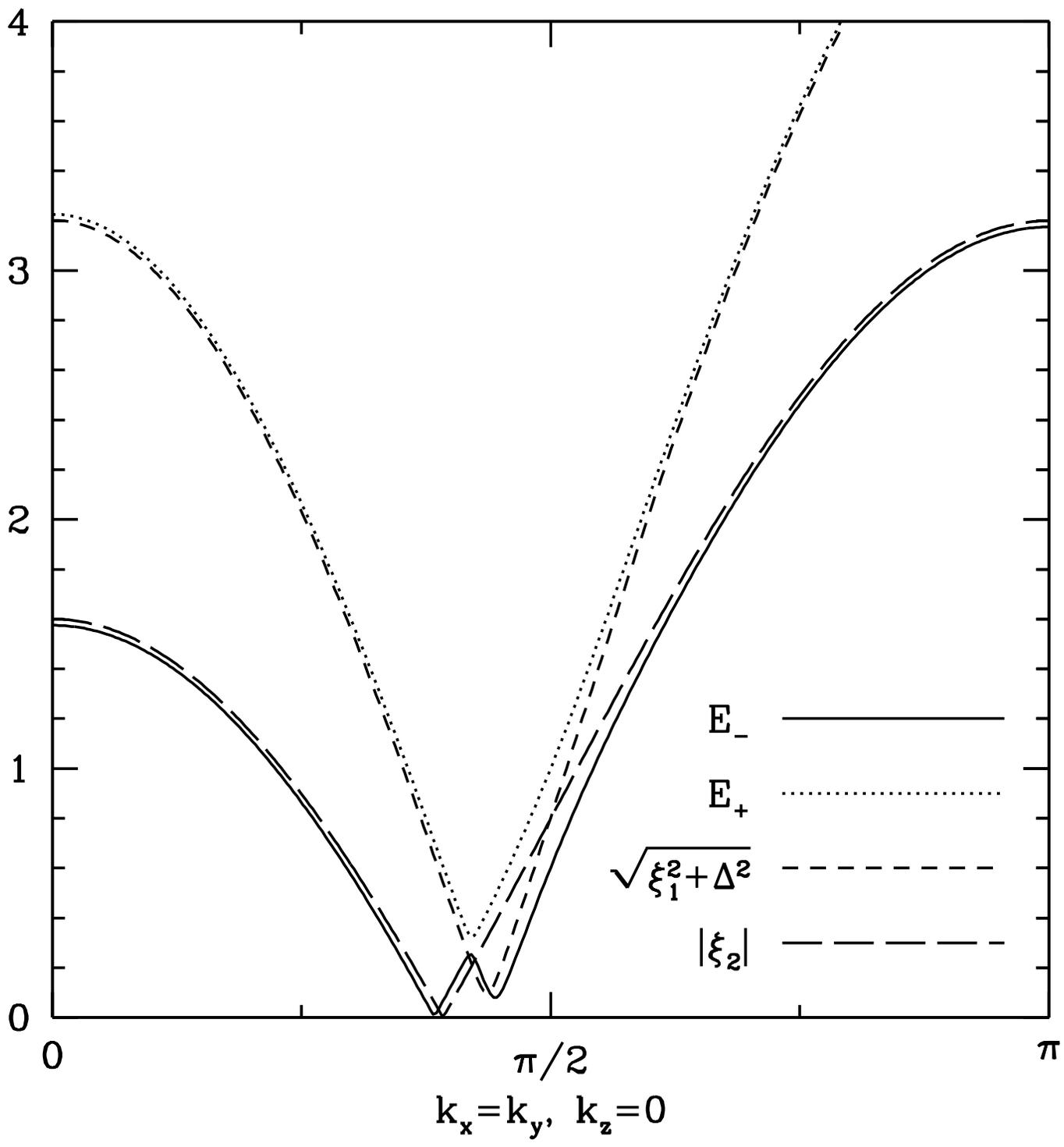
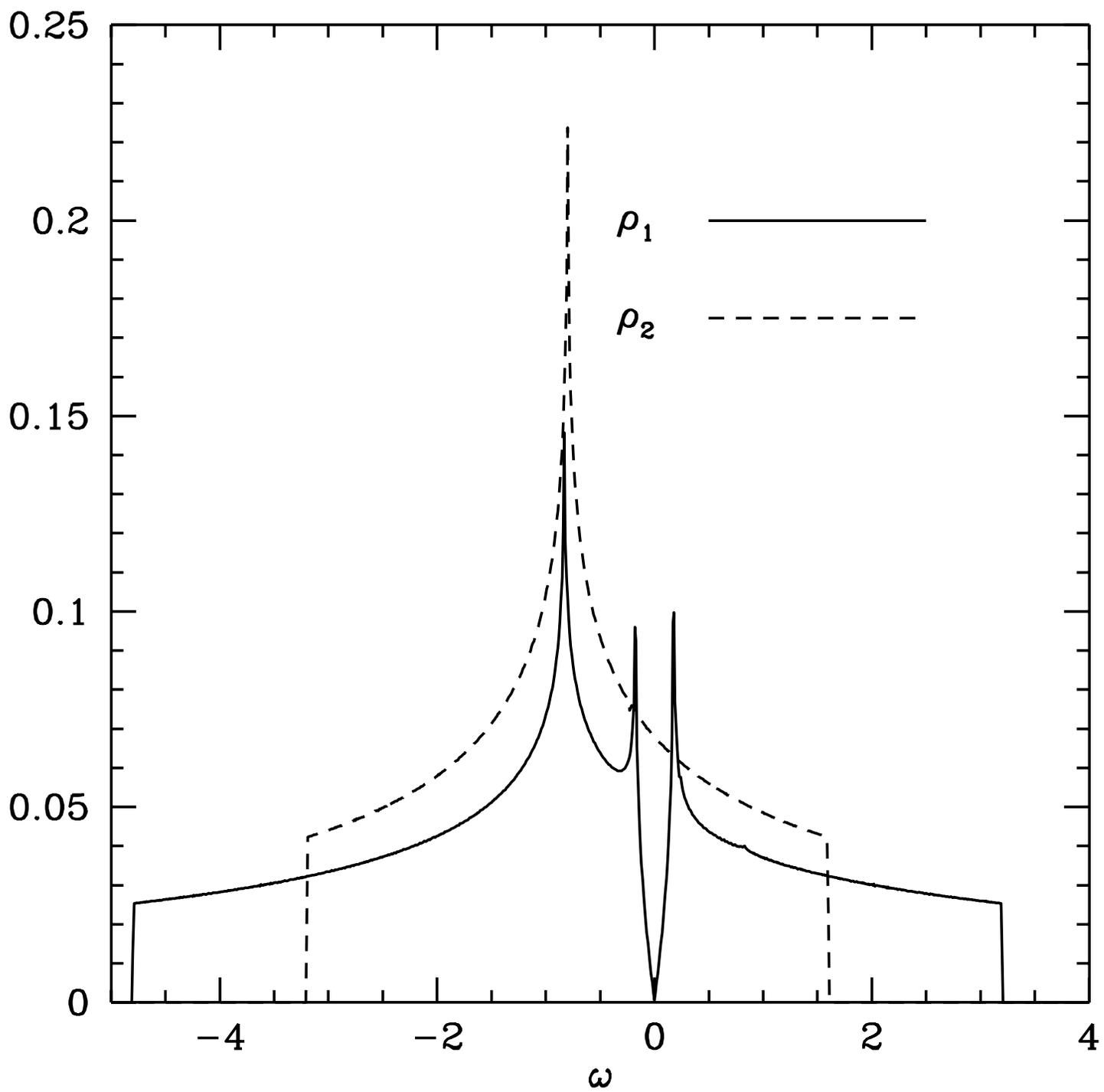
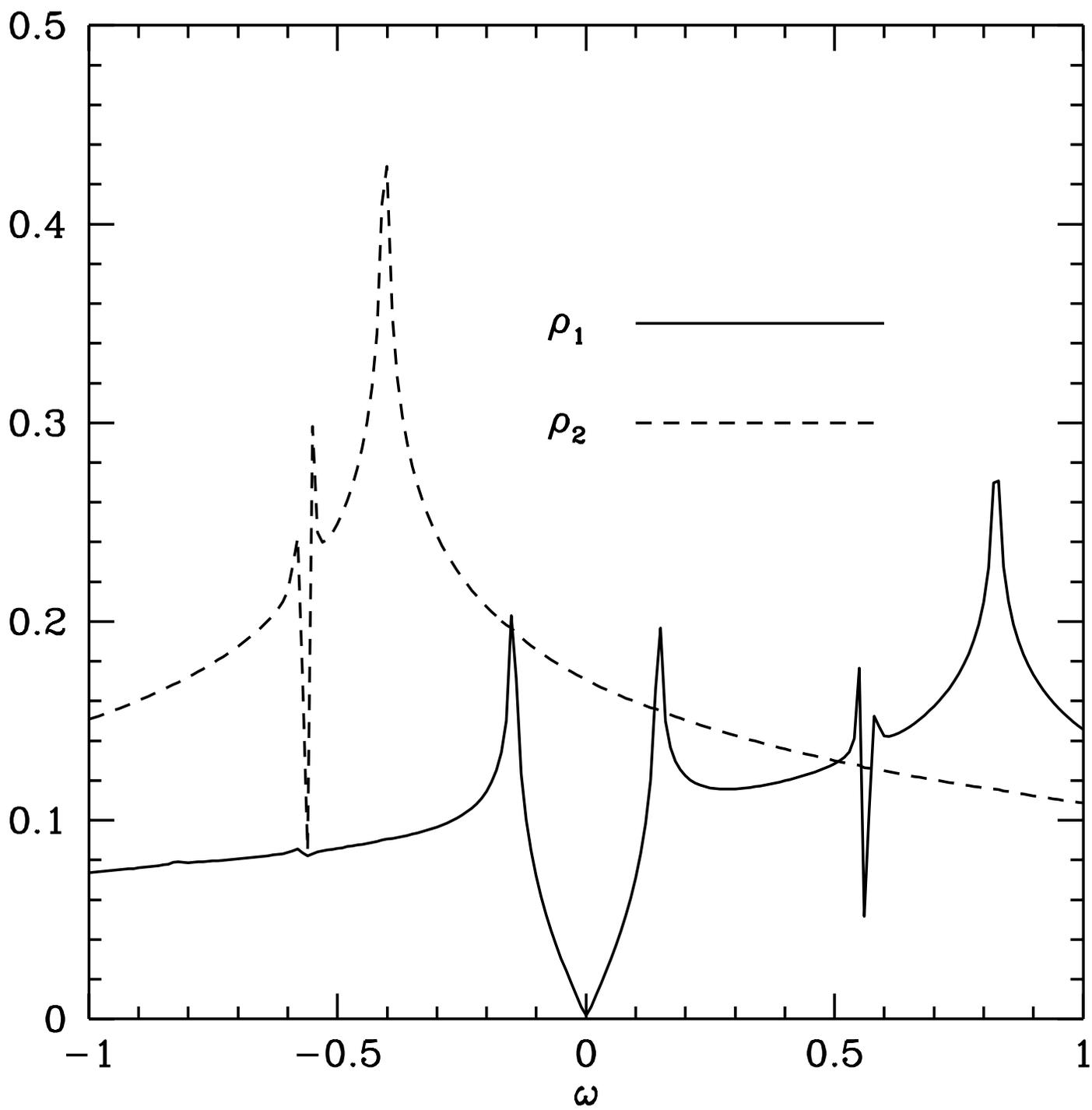
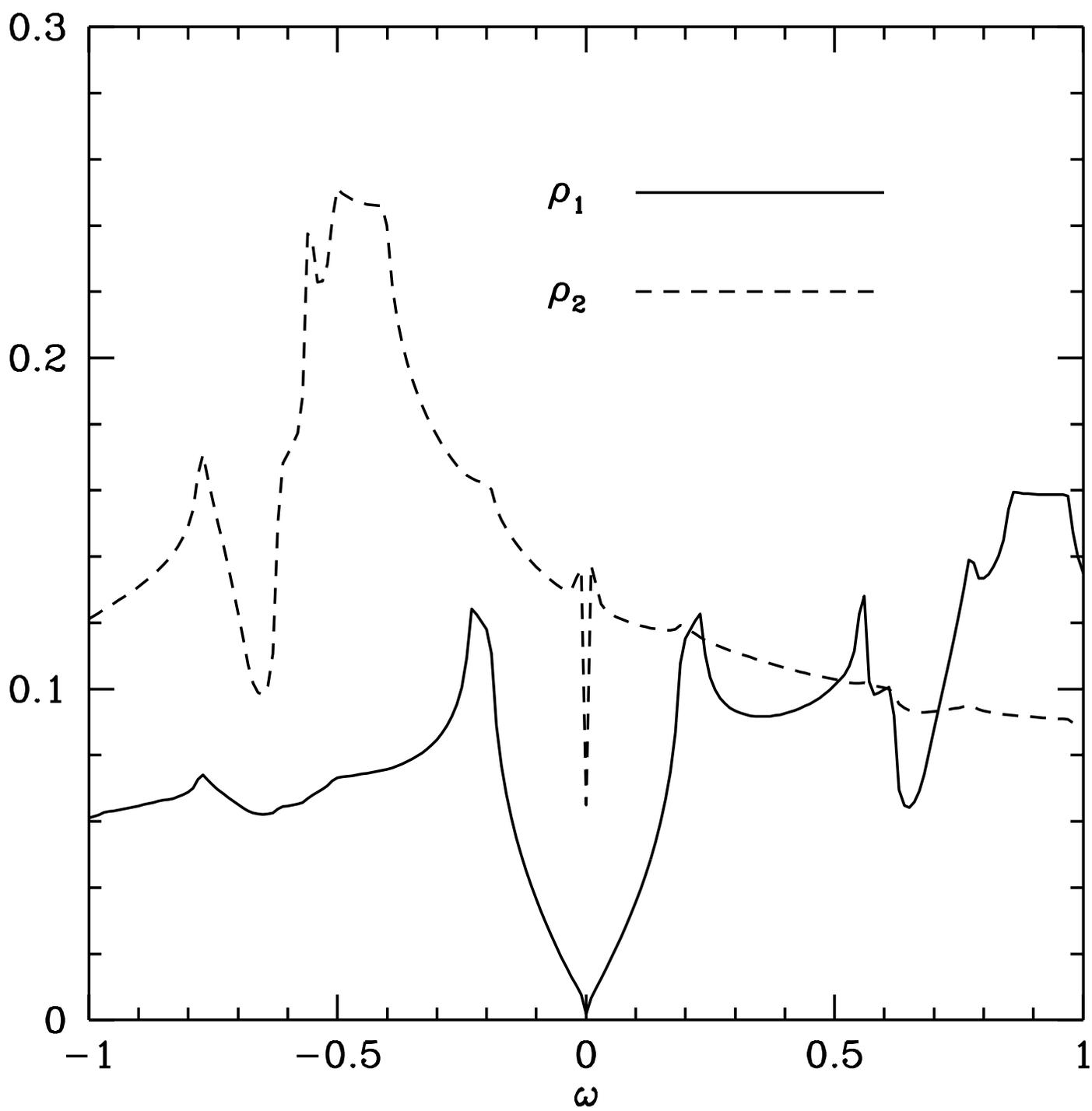
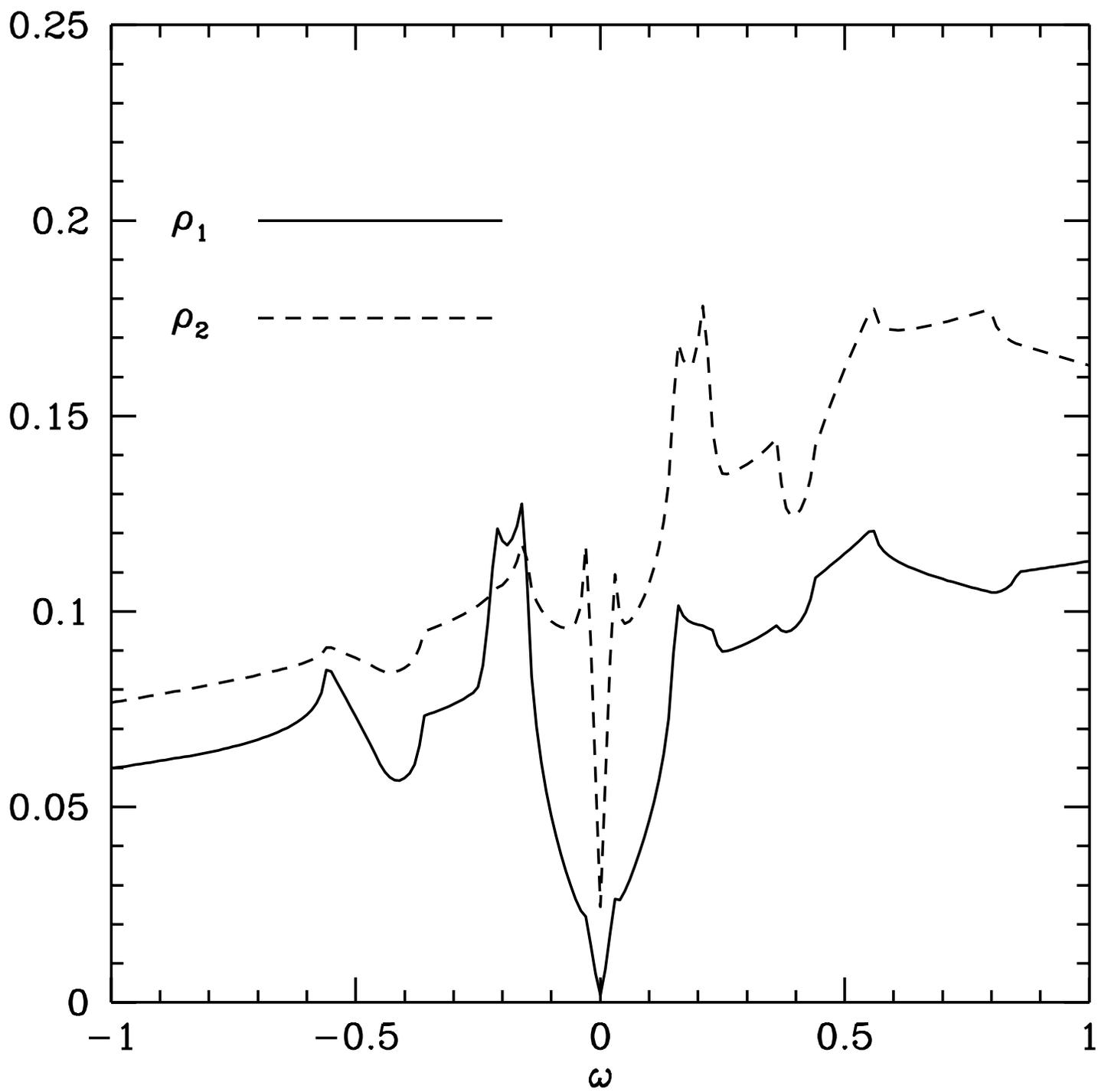
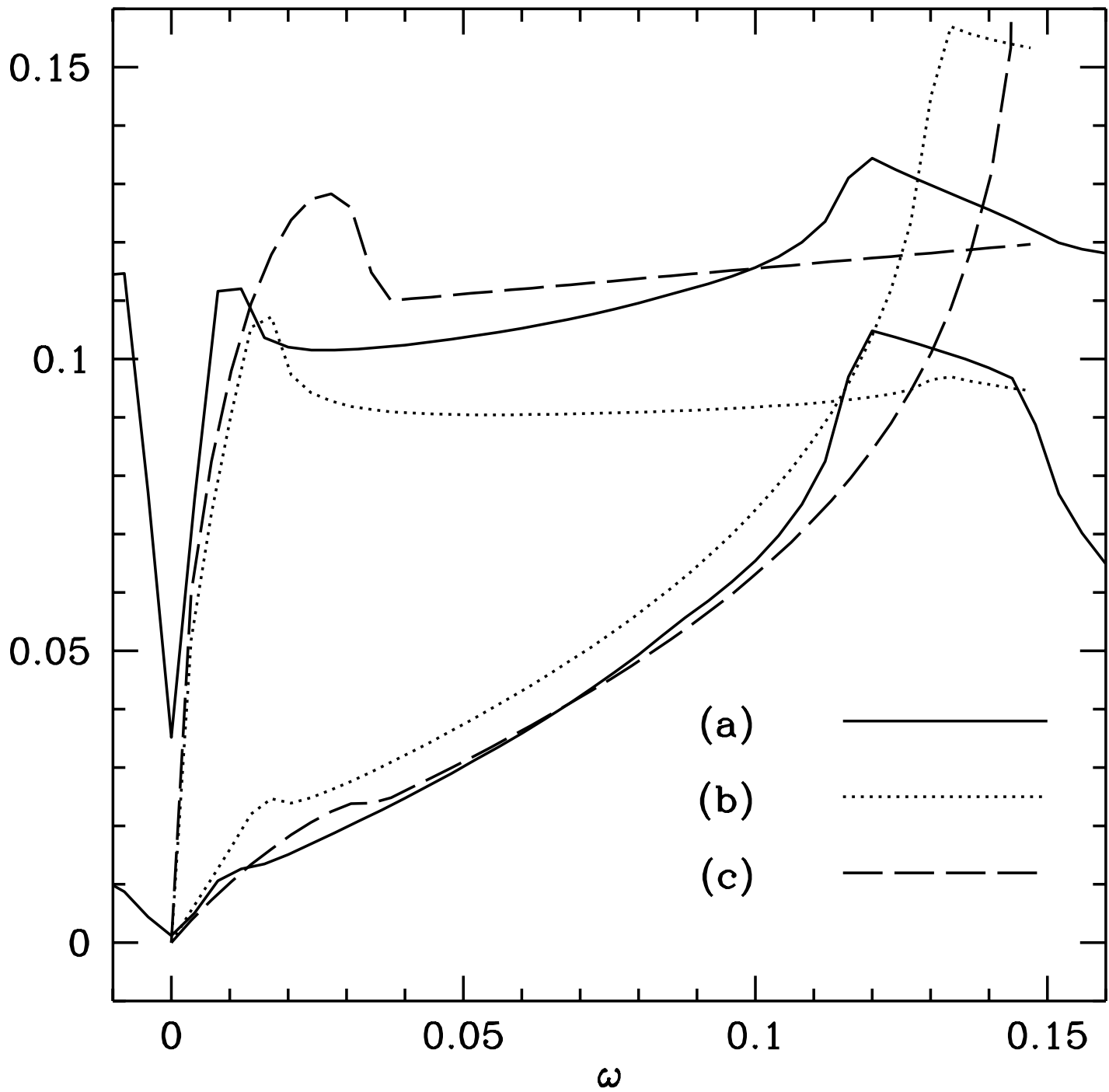
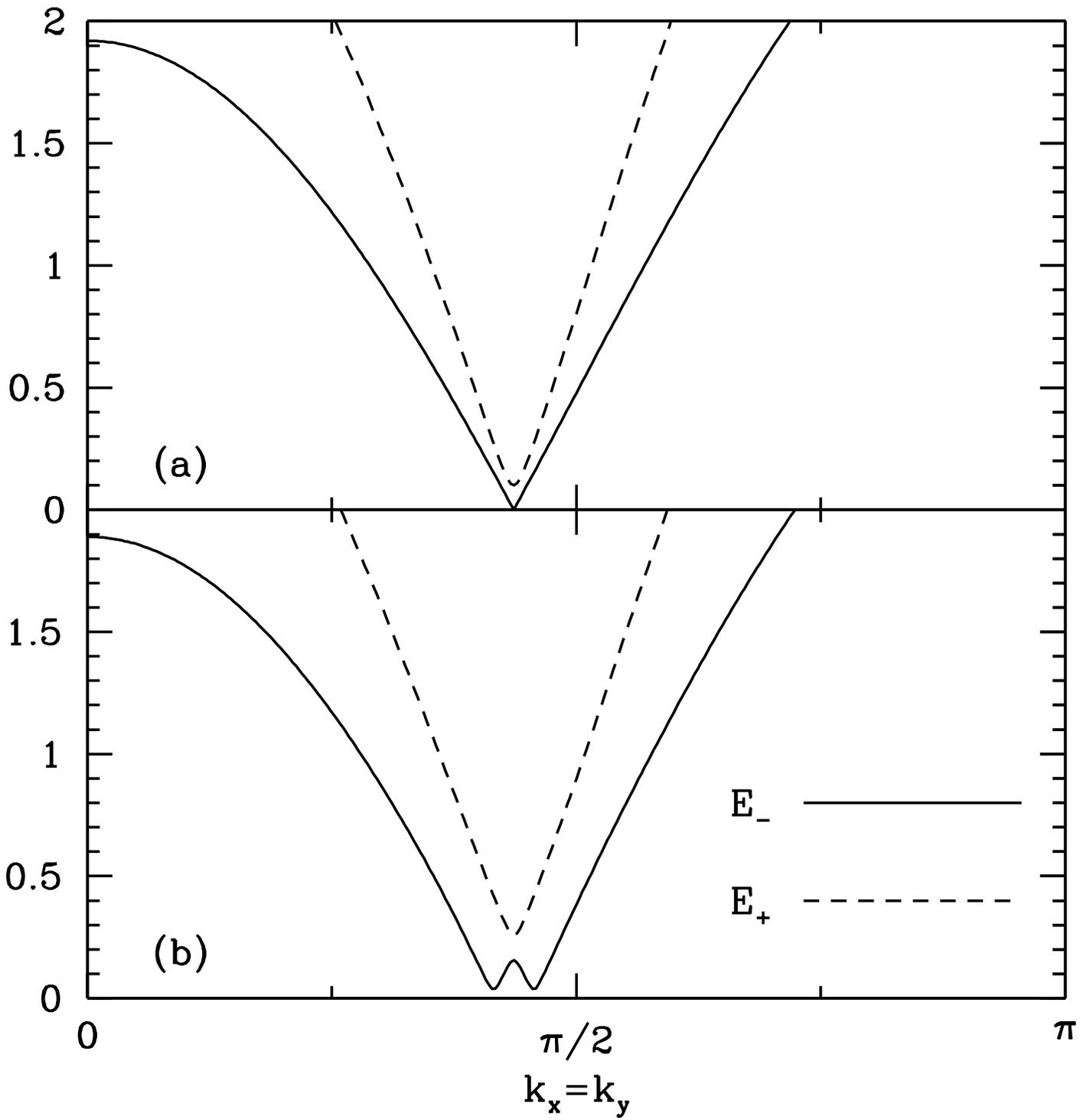
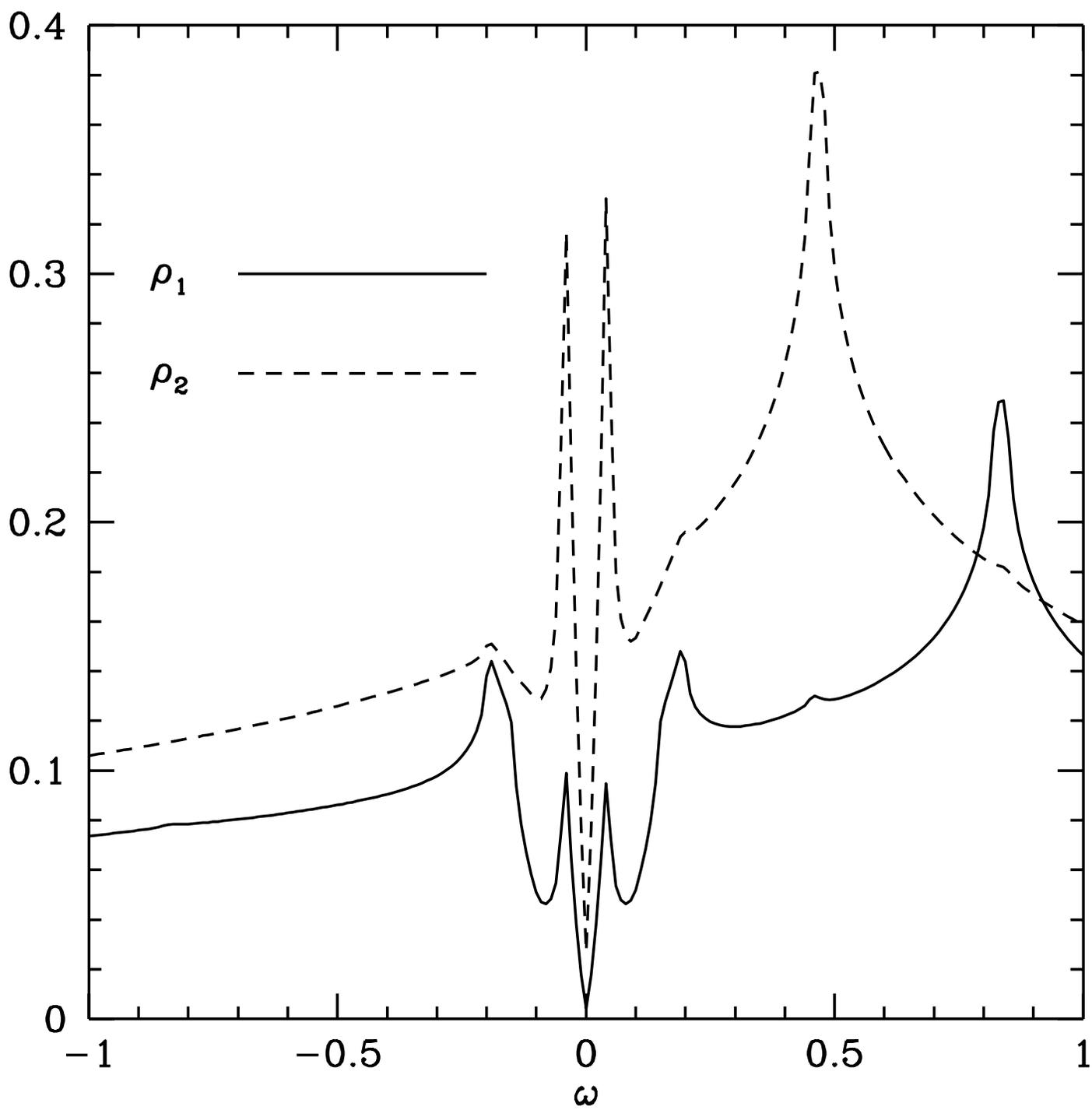
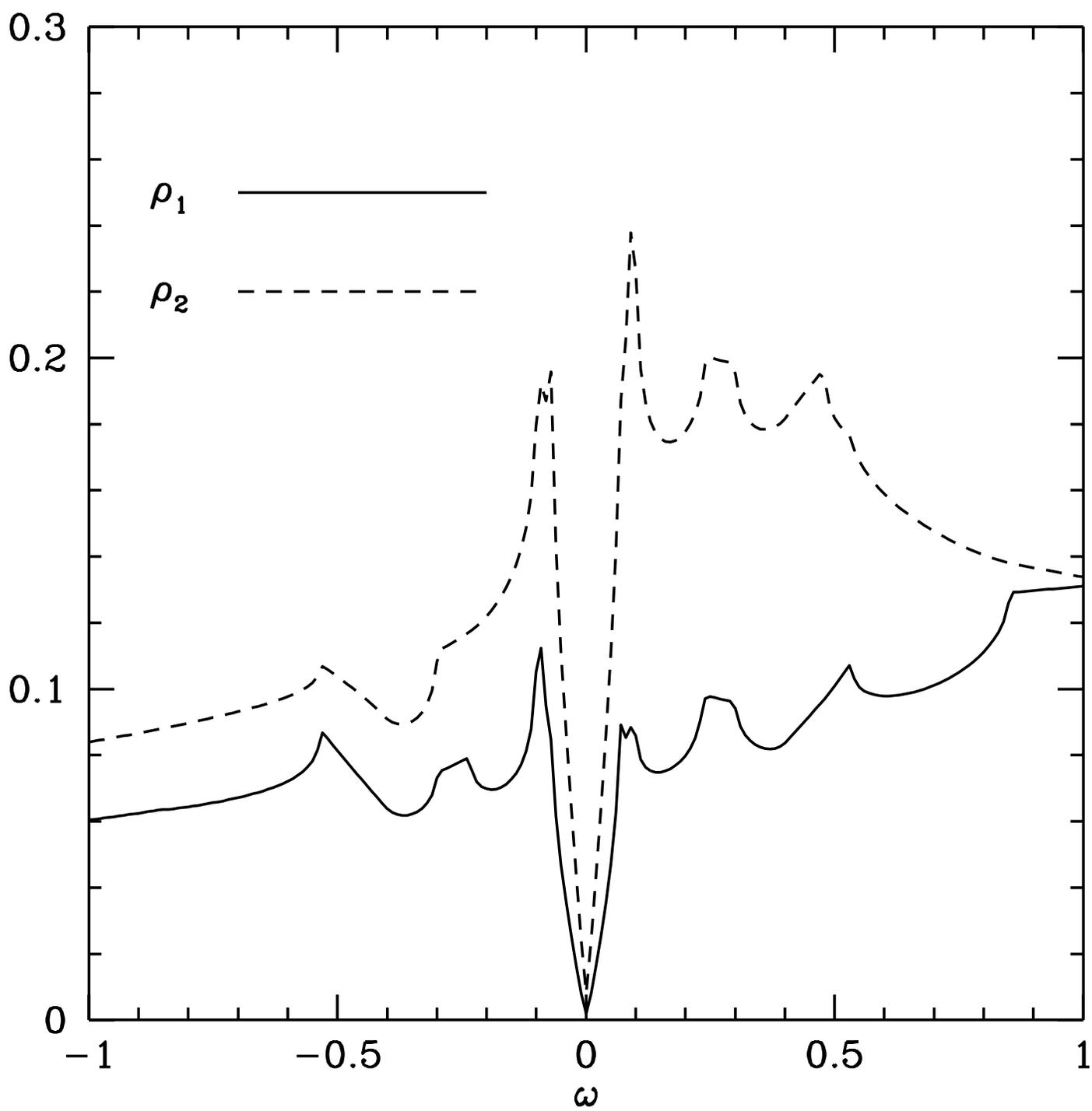
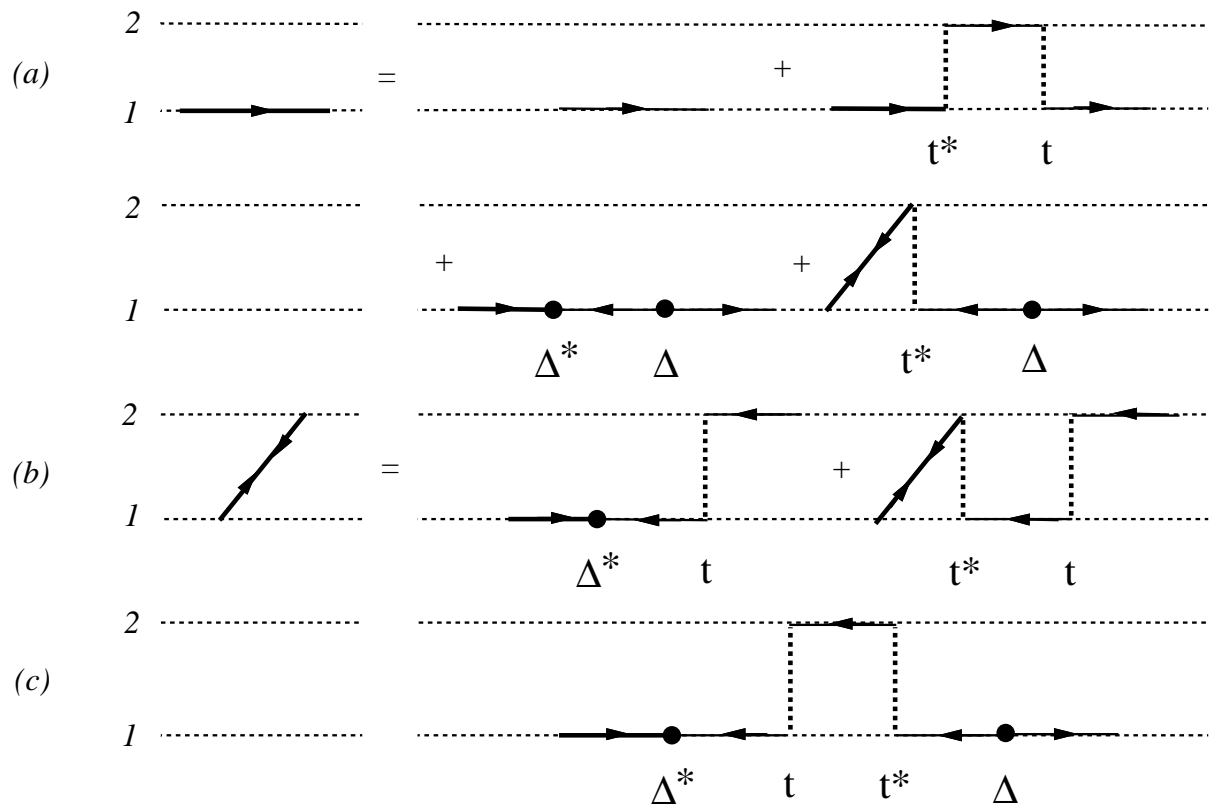
\end{document}